\documentclass[pre,aps,floats,twocolumn,superscriptaddress,floatfix]{revtex4n}
\usepackage{graphicx,amssymb,amsmath}

\begin{document}
%%%%%%%%%%%%%%%%%%%%%%%%%%%%%%%%%%%%%%%%%%%%%%%
\title{Thermodynamics of ideal quantum gas with fractional statistics in
  $\mathcal{D}$ dimensions} 
\author{Geoffrey G. Potter}
\affiliation{
  Department of Physics,
  University of Rhode Island,
  Kingston RI 02881, USA 
}
\author{Gerhard M\"uller}
\affiliation{
  Department of Physics,
  University of Rhode Island,
  Kingston RI 02881, USA 
}
\email[Gerhard M\"uller]{gmuller@uri.edu}
\author{Michael Karbach}
\affiliation{
  Bergische Universit{\"a}t Wuppertal,
  Fachbereich Mathematik und Naturwissenschaften,
  D-42097 Wuppertal, Germany}
\affiliation{
  Department of Physics,
  University of Rhode Island,
  Kingston RI 02881, USA 
}
\email[Michael Karbach]{michael@karbach.org}
%\centerline{(\footnotesize \version)}

%\ifthenelse{\equal{\writer}{gerhard}}%
%{\date{\today~--~1.0}} % Gerhard
%{\date{\version}}% Michael 
%\date{\today }
\date{\today}
%%%%%%%%%%%%%%%%%%%%%%%%%%%%%%%%%%%%%%%%%%%%%%%
\begin{abstract}
  We present exact and explicit results for the thermodynamic properties
  (isochores, isotherms, isobars, response functions, velocity of sound) of a
  quantum gas in dimensions $\mathcal{D}\geq1$ and with fractional exclusion
  statistics $0\leq g\leq1$ connecting bosons $(g=0)$ and fermions $(g=1)$. In
  $\mathcal{D}=1$ the results are equivalent to those of the Calogero-Sutherland
  model. Emphasis is given to the crossover between boson-like and fermion-like
  features, caused by aspects of the statistical interaction that mimic
  long-range attraction and short-range repulsion. The full isochoric heat
  capacity and the leading low-$T$ term of the isobaric expansivity in
  $\mathcal{D}=2$ are independent of $g$. The onset of Bose-Einstein
  condensation along the isobar occurs at a nonzero transition temperature in
  all dimensions. The $T$-dependence of the velocity of sound is in simple
  relation to isochores and isobars. The effects of soft container walls are
  accounted for rigorously for the case of a pure power-law potential.
\end{abstract}
\maketitle
%%%%%%%%%%%%%%%%%%%%%%%%%%%%%%%%%%%%%%%%%%%%%%%
%
\section{Introduction}\label{sec:intro}
%
%%%%%%%%%%%%%%%%%%%%%%%%%%%%%%%%%%%%%%%%%%%%%%%
A whole new arena for the study of effects of dimensionality has opened up in
the wake of the advances in instrumentation achieved in the research that led to
the observation of Bose-Einstein condensates \cite{PH02,PS03}. The magnetic and
optical traps that were developed and perfected in this and related lines of
research can also be made in shapes that effectively constrain the kinematics of
trapped gas atoms to fewer than $\mathcal{D}=3$ dimensions
\cite{GBM+01,GVL+01,SKC+01,PGS04}.  

In cigar-shaped traps, for example, the energy states of single particles with
nonzero momentum components perpendicular to the axis are frozen out for the
most part at low enough temperature.  In disk-shaped traps the only energy
states with significant occupancy at low $T$ are those with zero momentum
component perpendiclar to the plane. We are then dealing, effectively, with
quantum gases in dimensions $\mathcal{D}=1$ and $\mathcal{D}=2$, respectively. A
further effective change in spatial dimensionality can be produced by
controlling the firmness of the trap walls \cite{BPK87,BK91}.

Of no less importance than effects of dimensionality are, of course, effects of
interaction. They are notoriously difficult to handle in any effort that goes
beyond mean-field theory or low-order perturbation calculations. Here a little
explored non-perturbative approach presents itself as a promising alternative:
the thermodynamic analysis of statistically interacting quantum gases.
Statistical interaction grew out of the concept of fractional exclusion
statistics, introduced as a tool to interpret the quasiparticle composition of
the eigenstates in exactly solvable quantum many-body systems \cite{Hald91a}.

The thermodynamics of statistically interacting degrees of freedom is amenable
to exact analysis under very general circumstances \cite{Wu94}. A most
remarkable link exists between the \emph{dynamical} interaction in the form of
a coupling term in the many-body Hamiltonian and the \emph{statistical}
interaction in the form of a generalized Pauli principle.  For certain solvable
models in $\mathcal{D}=1$ one can replace the former by the latter and arrive
at equivalent thermodynamic properties \cite{BW94}.

The chances for extending any known exact solution for a dynamically interacting
quantum gas to $\mathcal{D}>1$ are slim because the criteria for purely
nondiffractive scattering \cite{Suth04} are unlikely to be satisfied apart from
highly contrived scenarios.  By contrast, extending the exact solution for a
statistically interacting quantum gas to higher dimensions is straightforward
conceptually if not technically.  

Notwithstanding the absence of an exact correspondence between dynamical and
statistical interactions in $\mathcal{D}>1$, the exact thermodynamics of
statistically interacting degrees of freedom has the potential of displaying a
richness in phenomena unrivaled by mean-field theory or by perturbative
approaches. Most importantly, it incorporates a full and consistent account of
fluctuations.

Here we consider a statistical interaction limited to particles with identical
momenta, in which case it reduces to a statistical exclusion condition. The two
parameters of the system explored here are the coefficient of exclusion
statistics, $0\leq g\leq1$, which spans a bridge between bosons $(g=0)$ and fermions
$(g=1)$, and the dimensionality $\mathcal{D}\geq1$. While this system does exhibit
the attributes typical for an ideal quantum gas, several of the highlighted
features are indicative of specific aspects of the statistical interaction
including features of long-range attraction and short-range repulsion. Selected
results are known from previous work \cite{Suth71a,NW94,IAMP96,JSSB96}.

A brief review of fractional exclusion statistics and its use in statistical
mechanics is found in Sec.~\ref{sec:statint}. The application of these
statistical mechanical tools to the model system under consideration is
described in Sec.~\ref{sec:qgwfs}. The exact thermodynamics of the
ideal quantum gas with fractional exclusion statistics in a
$\mathcal{D}$-dimensional box with rigid walls is presented in
Sec.~\ref{sec:gcs}. The impact of soft container walls as might be relevant in
the context of atomic traps is discussed in Sec.~\ref{sec:eoscw}. The main
results are assessed in Sec.~\ref{sec:concl} as benchmarks for corresponding
results pertaining to quantum gases with less constrained statistical
interactions as are forthcoming from work in progress \cite{PMK06b}.

%\pagebreak

%%%%%%%%%%%%%%%%%%%%%%%%%%%%%%%%%%%%%%%%%%%%%%%
%
\section{Fractional exclusion 
statistics}\label{sec:statint}  
%
%%%%%%%%%%%%%%%%%%%%%%%%%%%%%%%%%%%%%%%%%%%%%%%
The core relation of exclusion statistics is a generalized Pauli principle as introduced by Haldane
\cite{Hald91a}. It expresses how the number of states available to one particle is
affected by the presence of other particles:
\begin{equation}\label{eq:core}
  \Delta d_i = -g_{i}\Delta n_i,
\end{equation}
where the index $i$ enumerates distinct particle species (a flexible notion)
and the $g_{i}$ are coefficients of (fractional) exclusion statistics. For free
bosons we have $g_{i}=0$ and for free fermions $g_{i}=1$.  Integrating
Eq.~(\ref{eq:core}) yields the dimensionality of the \emph{one-particle}
Hilbert space in the presence of $n_i$ particles of species $i$:
\begin{equation}\label{eq:d1p}
  d_i = A_i -g_{i}(n_i-1),
\end{equation}
where $A_i$ are constants. The dimensionality of the \emph{many-particle}
Hilbert space is 
\begin{equation}\label{eq:prodWa}
  W=\prod_i \left( \begin{tabular}{c} $d_i+n_i-1$ \\ $n_i$ 
\end{tabular}\right),
\end{equation}
where each factor represents the simple combinatorial problem of
placing $n_i$ particles among $d_i+n_i-1$ distinct states available to
them. 

A system of (dynamically) free quasiparticles with fractional exclusion
statistics has two principal specifications: (i) a set of orbitals at energies
$\epsilon_i$ and (ii) a set of statistical exclusion coefficients $g_{i}$.  The
grand partition function is of the
form \cite{Wu94,Isak94}
\begin{equation}
  \label{eq:71}
Z=\prod_i\left[\frac{1+w_i}{w_i}\right],
\end{equation}
where the $w_i$ are determined by the nonlinear algebraic
equations
\begin{equation}
  \label{eq:54}
  \frac{\epsilon_i-\mu}{k_BT} = \ln(1+w_i)-g_{i}\ln\left(\frac{1+w_i}{w_i}\right).
\end{equation}
The energy level occupancies $\langle n_i\rangle$ are inferred from the $w_i$ as 
\begin{equation}
  \label{eq:1}
  \langle n_i\rangle = \frac{1}{w_i+g_i}.
\end{equation}

To illustrate the link between the combinatorial expression (\ref{eq:prodWa})
and the statistical mechanical expression (\ref{eq:71}) we consider an open
system of $M$ orbitals populated by (dynamically) free particles with (uniform) exclusion
coefficients $g>0$. The number of distinct $n$-particle configurations implied
by (\ref{eq:prodWa}) is
%\begin{subequations}
%  \label{eq:2}
%  \begin{align}
%    W_n(M) & = \binom{A+(1-g)(n-1)}{n} \\
%    & =\frac{\Gamma\big(A+n-g(n-1)\big)}{\Gamma(n+1)\Gamma\big(A-g(n-1)\big)},  
%  \end{align}
%\end{subequations}
\begin{eqnarray}
  \label{eq:2}
    W_n(M) &=& \binom{A+(1-g)(n-1)}{n} \nonumber \\
    &=& \frac{\Gamma\big(A+n-g(n-1)\big)}{\Gamma(n+1)\Gamma\big(A-g(n-1)\big)},  
  \end{eqnarray}
where $A$ is to be adjusted so as to yield integer values for
$W_n(M)$ and a meaningful maximum capacity $n_\mathrm{max}$.  For integer $g$
we use $A=M$. For $g=\frac{1}{2}$ we use $A=M$ if $M$ is odd and
$A=M+\frac{1}{2}$ if $M$ is even. This scheme is naturally extended to other
fractional values of $g$, limiting the maximum capacity to
$n_\mathrm{max}<(M+g)/g$.  

The dimensionality of the Fock space,
\begin{equation}
  \label{eq:3}
  W_\mathrm{tot}(M)=\sum_{n=0}^{n_\mathrm{max}}W_n(M),
\end{equation}
can be expressed, for integer $g$, as a (higher-order) Lam{\'e} sequence,
% \begin{align}
%   \label{eq:5}
%   W_\mathrm{tot}(M) \!=\! 
%   \begin{cases}
%     M+1 & : M=0,1,\!\ldots\!,g\!-\!1,\\    
%     W_\mathrm{tot}(M\!-\!1)\!+\!W_\mathrm{tot}(M\!-\!g) &: M\!=\!g,g\!+\!1,\ldots
%   \end{cases}
% \end{align}
\begin{equation}
  \label{eq:6aa}
  W_\mathrm{tot}(M) = M+1 
\end{equation}
for $M=0,1,\ldots,g-1$ and 
\begin{equation}
  \label{eq:6bb}
   W_\mathrm{tot}(M) = W_\mathrm{tot}(M-1)+W_\mathrm{tot}(M-g)
 \end{equation}
 for $M=g,g+1,\ldots$ In the case of fermions $(g=1)$ we obtain
 $W_\mathrm{tot}(M)=2^M$ and for $g=2$ a string of Fibonacci numbers:
 $W_\mathrm{tot}(M)=F_M=1,2,3,5,8,\ldots$ For semions $(g=\frac{1}{2}$) we use
 (\ref{eq:2}). The result is a string of alternate Fibonacci numbers:
 $W_\mathrm{tot}(M)=F_{2M+1}=1,3,8,\ldots$

The quantity we investigate for our illustration is the limit $M\to\infty$ of the
proportion by which the number of many-particle states increases if we add one
orbital to the system: $R_M\doteq W_\mathrm{tot}(M+1)/W_\mathrm{tot}(M)$. For
fermions we have $R_M=2$ independent of $M$ as expected. For semions we obtain
the square of the golden section:
\begin{equation}
  \label{eq:7}
  R_M=\frac{F_{2M+3}}{F_{2M+1}}~
  \stackrel{M\to\infty}{\longrightarrow}~ \left[\frac{1}{2}\left(\sqrt{5}+1\right)\right]^2. 
\end{equation}
Now we derive $R_\infty$ directly from (\ref{eq:71}) and (\ref{eq:54}) at $T=\infty$:
\begin{equation}
  \label{eq:9}
  R_\infty=\frac{1+w}{w},\quad w^g(1+w)^{1-g}=1.
\end{equation}
The solution for fermions is $w=1$, yielding $R_\infty=2$ as before via
combinatorics. In the semion case the asymptotic result in (\ref{eq:7}) is
recovered from the solution of the quadratic equation in (\ref{eq:9}).

%%%%%%%%%%%%%%%%%%%%%%%%%%%%%%%%%%%%%%%%%%%%%%%
%
\section{Quantum gas with fractional 
statistics}\label{sec:qgwfs}  
%
%%%%%%%%%%%%%%%%%%%%%%%%%%%%%%%%%%%%%%%%%%%%%%%

Before we go on to discuss how Eqs. (\ref{eq:71})--(\ref{eq:1}) are applied to
a quantum gas in $\mathcal{D}$ dimensions, we turn to an exactly solved model
in $\mathcal{D}=1$.

%%%%%%%%%%%%%%%%%%%%%%%%%%%%%%%%%%%%%%%%%%%%%%%
%
\subsection{Calogero-Sutherland model}\label{sec:etdi}  
%
%%%%%%%%%%%%%%%%%%%%%%%%%%%%%%%%%%%%%%%%%%%%%%%
The Calogero-Sutherland (CS) model describes massive particles on a ring
of circumference $L$ with a (dynamical) pair interaction (in units where $\hbar^2/2m=1$)
\cite{Calo71,Suth71,Suth71a,Suth72}:
\begin{equation}
  \label{eq:77cs}
  H = -\sum_{i=1}^N\frac{\partial^2}{\partial x_i^2}+ \sum_{j<i}\frac{2g(g-1)}{d_{ij}^2}.
\end{equation}
Here
\begin{equation}
  \label{eq:26}
d_{ij} = \left|\frac{L}{\pi}\sin\left(\frac{\pi}{L}(x_i-x_j)\right)\right|
\end{equation}
represents the chord distance on the ring between particles.
Varying the parameter $g$ across the interval $0\leq g \leq1$, amounts to a
parametric link between free bosons $(g=0)$ and free fermions $(g=1)$.

The spectrum and thermodynamics of the CS model can be described by an
asymptotic coordinate Bethe ansatz \cite{Suth04}.  The thermodynamic Bethe
ansatz inferred therefrom expresses the grand potential in the form
\begin{equation}
  \label{eq:73}
  \Omega(T,L,\mu)=-k_BT\left(\frac{L}{2\pi}\right)\int_{-\infty}^{+\infty}dk\ln\left(1+e^{-\epsilon(k)/k_BT}\right),
\end{equation}
where $\epsilon(k)$ is determined by the Yang-Yang-type equation \cite{YY69}
%\begin{equation}
%  \label{eq:74}
%\epsilon(k)=k^2-\mu-\frac{k_BT}{2\pi}\int_{-\infty}^{+\infty}dk'K(k-k')\ln\left(1+e^{-\epsilon(k')/k_BT}\right)
%\end{equation}
\begin{eqnarray}
  \label{eq:74}
\epsilon(k) &=& k^2-\mu \nonumber \\
&&\hspace*{-12mm} -\frac{k_BT}{2\pi}\int_{-\infty}^{+\infty}dk'K(k-k')\ln\left(1+e^{-\epsilon(k')/k_BT}\right)
\end{eqnarray}
with kernel
\begin{equation}
  \label{eq:79}
  K(k-k')=2\pi(1-g)\delta(k-k').
\end{equation}
The distribution of particles, $\rho_P(k)$, is found to be the
solution, for given $\epsilon(k)$, of the Lieb-Liniger-type equation \cite{LL63}
\begin{equation}
  \label{eq:4}
  2\pi\rho_P(k)\left[1+e^{\epsilon(k)/k_BT}\right] = 1+\int_{-\infty}^{+\infty}dk'\,K(k-k')\rho_P(k').  
\end{equation}

%\pagebreak

%%%%%%%%%%%%%%%%%%%%%%%%%%%%%%%%%%%%%%%%%%%%%%%
%
\subsection{Generalization of CS model}\label{sec:gocs}  
%
%%%%%%%%%%%%%%%%%%%%%%%%%%%%%%%%%%%%%%%%%%%%%%%
Returning to statistical interaction we consider a nonrelativistic gas in a
rigid box of dimensionality $\mathcal{D}$ and volume $V=L^\mathcal{D}$ with
energy-momentum relation $\epsilon_0(k)=|\mathbf{k}^2|$ and exclusion statistics
\begin{equation}
  \label{eq:19}
  g(\mathbf{k}-\mathbf{k}')=g\delta(\mathbf{k}-\mathbf{k}').
\end{equation}
The grand potential derived from (\ref{eq:71}) becomes
\begin{equation}
  \label{eq:12}
  \Omega= -k_BT\left(\frac{L}{2\pi}\right)^\mathcal{D} \int
  d^\mathcal{D}k\,\ln\frac{1+w_\mathbf{k}}{w_{\mathbf{k}}},
\end{equation}
where $w_\mathbf{k}$ is the solution of 
\begin{equation}
  \label{eq:13}
  \frac{|\mathbf{k}|^2-\mu}{k_BT}= \ln(1+w_\mathbf{k}) 
  -g\ln\frac{1+w_{\mathbf{k}}}{w_{\mathbf{k}}}.
\end{equation}
The particle density in $\mathbf{k}$-space is represented by the function 
\begin{equation}
  \label{eq:14}
  \langle n_{\mathbf{k}}\rangle =\frac{1}{w_{\mathbf{k}}+g}.
\end{equation}
The fundamental thermodynamic relations, from which most thermodynamic
properties are conveniently derived, are integrals involving $w_\mathbf{k}$ and
$\langle n_\mathbf{k}\rangle$:
\begin{equation}
  \label{eq:15}
  \frac{pV}{k_BT} = \left(\frac{L}{2\pi}\right)^\mathcal{D} \int
  d^\mathcal{D}k\,\ln\frac{1+w_\mathbf{k}}{w_{\mathbf{k}}},
\end{equation}
\begin{equation}
  \label{eq:16}
  \mathcal{N}= \left(\frac{L}{2\pi}\right)^\mathcal{D} \int
  d^\mathcal{D}k\, \langle n_{\mathbf{k}}\rangle,
\end{equation}
\begin{equation}
  \label{eq:17}
 U= \left(\frac{L}{2\pi}\right)^\mathcal{D} \int
  d^\mathcal{D}k\, |\mathbf{k}|^2\langle n_{\mathbf{k}}\rangle.
\end{equation}
These relations state the dependence of the pressure $p$, the average number of
particles $\mathcal{N}$, and the internal energy $U$ on fugacity $z$,
temperature $T$, and volume $V$. They are also known as the \emph{thermodynamic}
equation of state (\ref{eq:15})-(\ref{eq:16}) and the \emph{caloric} equation of
state (\ref{eq:16})-(\ref{eq:17}).

It was recognized \cite{BW94,Isak94,MS99} that the thermodynamic Bethe ansatz solution
(\ref{eq:73})--(\ref{eq:4}) of the CS model
is equivalent
to the solution of a statistical interacting gas as presented above for
$\mathcal{D}=1$ if we make the following identifications:
\begin{equation}
  \label{eq:75}
  w_k=\exp\left(\frac{\epsilon(k)}{k_BT}\right),\quad  \langle n_k\rangle= 2\pi\rho_P(k),
\end{equation}
\begin{equation}
  \label{eq:76}
  g(k,k')=\delta(k-k')-\frac{1}{2\pi}K(k-k').
\end{equation}
For the thermodynamic extension to $\mathcal{D}\geq1$ of the CS model we use the
statistical interaction (\ref{eq:19}) and the density of 1-particle states of
the nonrelativistic ideal gas,
\begin{equation}
  \label{eq:22}
  D_0(\epsilon_0)=\left(\frac{L}{2\pi}\right)^\mathcal{D}
  \frac{\pi^{\mathcal{D}/2}}{\Gamma(\mathcal{D}/2)} \,\epsilon_0^{\mathcal{D}/2-1},
\end{equation}
to rewrite
Eqs.~(\ref{eq:12})--(\ref{eq:14}) in the form
\begin{equation}
  \label{eq:23}
  \Omega= -k_BT\int_0^\infty d\epsilon_0\,D_0(\epsilon_0)\ln\frac{1+w(\epsilon_0)}{w(\epsilon_0)},
\end{equation}
\begin{subequations}
  \label{eq:24}
  \begin{align}
    \frac{1}{z}\exp\left(\frac{\epsilon_0}{k_BT}\right) & =
    [w(\epsilon_0)]^g[1+w(\epsilon_0)]^{1-g},
    \\
    z & = \exp(\mu/k_BT),
\end{align}
  
\end{subequations}
\begin{equation}
  \label{eq:25}
  \langle n(\epsilon_0)\rangle=\frac{1}{w(\epsilon_0)+g},
\end{equation}
and to reduce Eqs. (\ref{eq:15})--(\ref{eq:17}) to integrals over functions of $\epsilon_0$.

%%%%%%%%%%%%%%%%%%%%%%%%%%%%%%%%%%%%%%%%%%%%%%%
%
\subsection{Average level occupancy}\label{sec:alo}  
\begin{figure}[htb]
\centering \includegraphics[width=80mm]{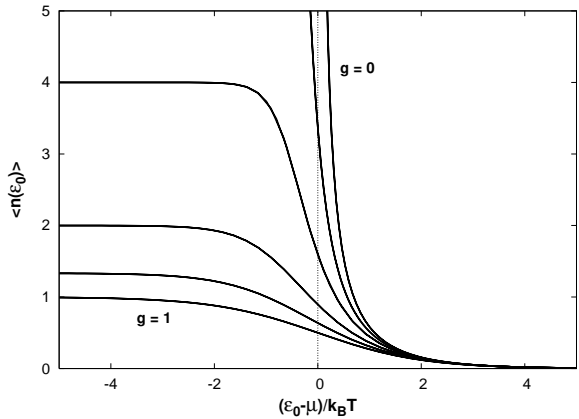}
  \caption{Average level occupancy $\langle n(\epsilon_0)\rangle$ versus $(\epsilon_0-\mu)/k_BT$ for
    $g=0$, $0.1$, $0.25$, $0.5$, $0.75$, $1$ (from top down).}\label{fig:csalo} 
\end{figure}

%
%%%%%%%%%%%%%%%%%%%%%%%%%%%%%%%%%%%%%%%%%%%%%%%
In ideal quantum gases the average level occupancy $\langle n(\epsilon_0)\rangle$ is a unique
function of $(\epsilon_0-\mu)/k_BT$ independent of $\mathcal{D}$. A plot of this function for
the generalized CS model is shown in Fig.~\ref{fig:csalo} for various
values of $g$. In the boson case we have $w(\epsilon_0)=e^{(\epsilon_0-\mu)/k_BT}-1$ and
in the fermion case $w(\epsilon_0)=e^{(\epsilon_0-\mu)/k_BT}$. For semions
we have
\begin{equation}
  \label{eq:58}
w(\epsilon_0)  = \sqrt{\left[e^{(\epsilon_0-\mu)/k_BT}\right]^2 +\frac{1}{4}}
  -\frac{1}{2}.
\end{equation}
The exact analytic solution of (\ref{eq:24}) with $0<g<1$ is known in a series
representation \cite{JSSB96}. The asymptotic level occupancy for $(\epsilon_0-\mu)/k_BT\to-\infty$ is $1/g$.
%%%%%%%%%%%%%%%%%%%%%%%%%%%%%%%%%%%%%%%%%%%%
%%%%%%%%%%%%%%%%%%%%%%%%%%%%%%%%%%%%%%%%%%%%%

%\pagebreak

%%%%%%%%%%%%%%%%%%%%%%%%%%%%%%%%%%%%%%%%%%%%%%%
%
\section{Thermodynamics of the 
generalized CS model}\label{sec:gcs}  
%
%%%%%%%%%%%%%%%%%%%%%%%%%%%%%%%%%%%%%%%%%%%%%%%
Exact and explicit results for the thermodynamics of the generalized CS model
can now be calculated from the general expressions derived in
Sec.~\ref{sec:qgwfs}.

%%%%%%%%%%%%%%%%%%%%%%%%%%%%%%%%%%%%%%%%%%%%%%%
%
\subsection{CS functions}\label{sec:csf}  
%
%%%%%%%%%%%%%%%%%%%%%%%%%%%%%%%%%%%%%%%%%%%%%%%
Introducing the CS functions
\begin{align}
  \label{eq:53}
  & G_{n}(z,g) \doteq \frac{1}{\Gamma(n)} \int_{0}^{\infty} dx\; 
  \frac{x^{n-1}}{\bar{w}(x)+g}, \\
   & [\bar{w}(x)]^g[1+\bar{w}(x)]^{1-g}=\frac{e^x}{z},
\end{align}
for $n>0$
we rewrite Eqs.~(\ref{eq:15})--(\ref{eq:17}) as
\begin{equation}
  \label{eq:50}
  \frac{p\lambda_T^\mathcal{D}}{k_BT}=G_{\mathcal{D}/2+1}(z,g),
\end{equation}
\begin{equation}
  \label{eq:52}
  \frac{\mathcal{N}\lambda_T^\mathcal{D}}{V}=G_{\mathcal{D}/2}(z,g)\quad 
\left[ +\frac{z}{1-z}\right],
\end{equation}
\begin{equation}
  \label{eq:51} 
  \frac{U\lambda_T^\mathcal{D}}{V}\left/ \left(\frac{\mathcal{D}}{2}\,k_BT\right) \right.
=G_{\mathcal{D}/2+1}(z,g),  
\end{equation}
where 
\begin{equation}
  \label{eq:10}
  \lambda_T\doteq\sqrt{\frac{h^2}{2\pi mk_BT}} ~\stackrel{\hbar^2/2m=1}{\longrightarrow}~ \sqrt{\frac{4\pi}{k_BT}}
\end{equation}
is the thermal wavelength and the term in \eqref{eq:52} enclosed by brackets is
relevant only if $g=0$ and $\mathcal{D}>2$. For $g=0$ and $g=1$ the CS functions
become the Bose-Einstein (BE) functions and Fermi-Dirac (FD) functions,
respectively. The range of fugacity is $0\leq z\leq1$ for bosons. For all other cases
$z$ has no upper bound.

\begin{figure}[hbt]
  \centering
  \includegraphics[width=80mm]{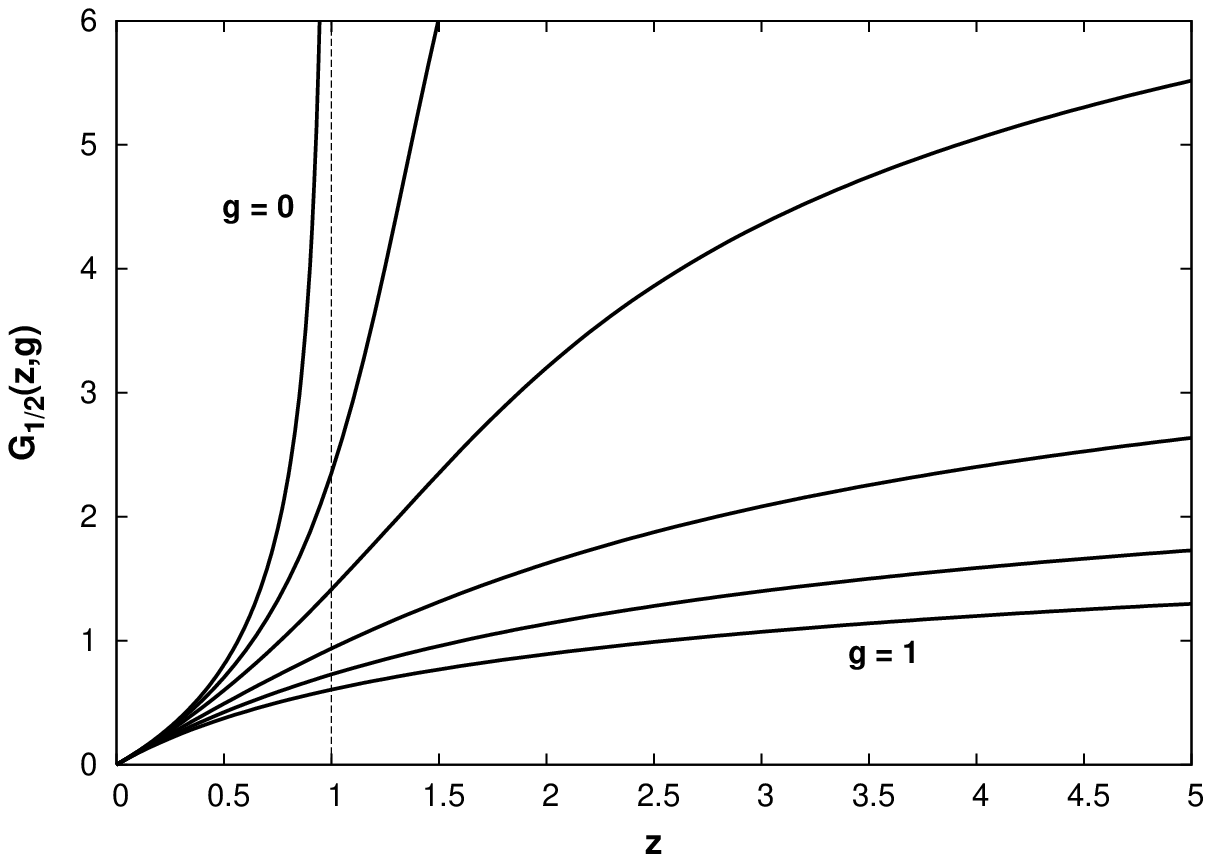}
%\end{figure}
%\begin{figure}[hbt]
%  \centering 
  \includegraphics[width=80mm]{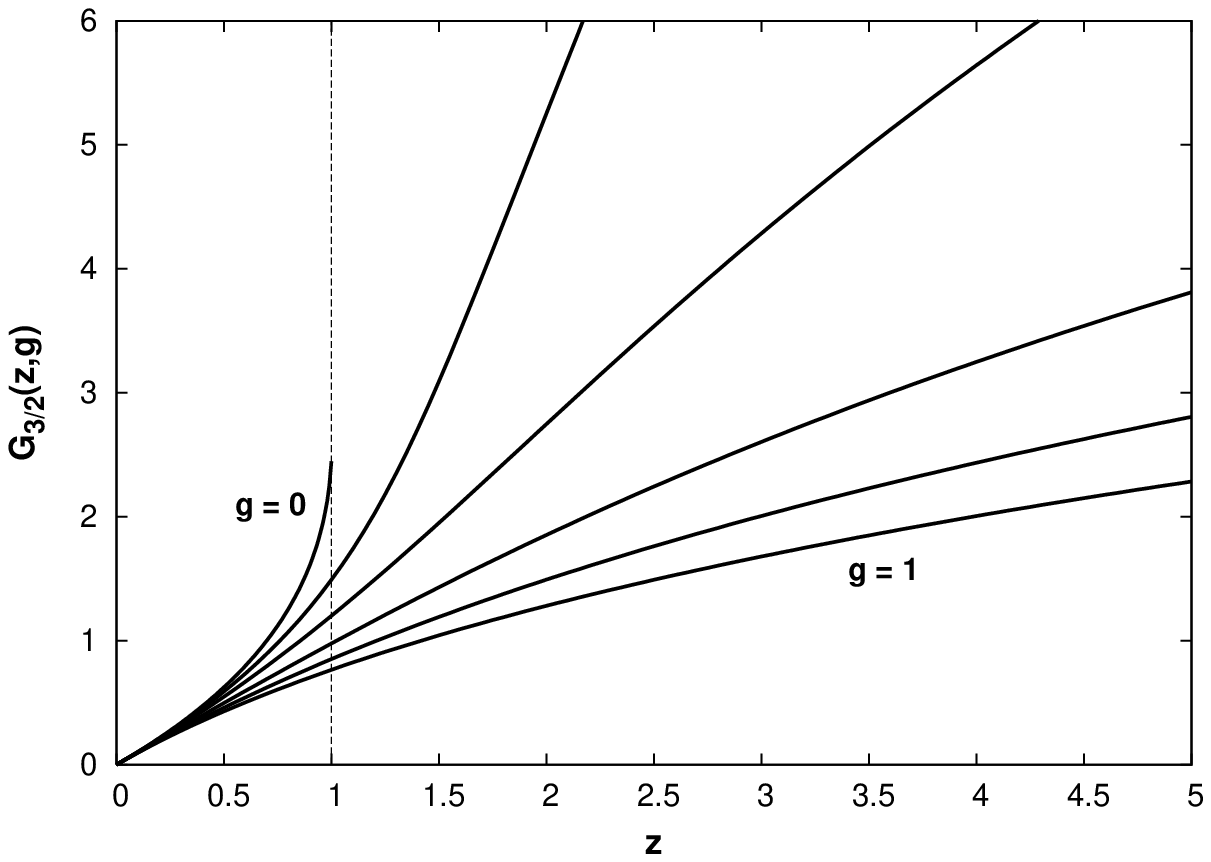}
  \caption{CS functions (\ref{eq:53}) for $n=\frac{1}{2},\frac{3}{2}$
    plotted versus $z$ for $g = 0$, $0.1$, $0.25$, $0.5$, $0.75$ (from top down).}\label{fig:Gnzotf}
\end{figure}

The overall shape of the CS functions is illustrated in Fig.~\ref{fig:Gnzotf}.
For $\frac{1}{2}\leq g\leq1$ the curves are concave. For $0<g\leq\frac{1}{2}$ they have a
convex portion at small $z$ and then switch to to concave behavior. The curve
for $g=0$ is convex over its entire (restricted) domain.

%%%%%%%%%%%%%%%%%%%%%%%%%%%%%%%%%%%%%%%%%%%%
%%%%%%%%%%%%%%%%%%%%%%%%%%%%%%%%%%%%%%%%%%%%%

The CS functions have the following power series expansions \cite{JSSB96}:
\begin{align}
  \label{eq:114}
  G_n(z,g) &= \sum_{l=1}^\infty \frac{z^l}{l^n}\frac{\Gamma(l-lg)}{\Gamma(l)\Gamma(1-lg)}
  \nonumber \\
  &= z+\frac{z^2}{2^n}(1\!-\!2g)
  +\frac{z^3}{3^n}(1\!-\!3g)\left(1\!-\!\frac{3}{2}g\right)+\mathcal{O}(z^4). 
\end{align}
The radius of convergence depends on $g$:
\begin{align}
  r(g) = \frac{1}{g^{g}(1-g)^{1-g}}.
\end{align}
The first two terms of the asymptotic expansion for large $z$ at $g>0$ are
\cite{JSSB96,IAMP96} 
\begin{equation}
  \label{eq:100}
  G_n(z,g) \stackrel{z\to\infty}{\leadsto} 
  \frac{(\ln z)^n}{g\Gamma(n+1)}\left[1+ \frac{\pi^2}{6}\, \frac{gn(n-1)}{(\ln z)^{2}}+\ldots
  \right].
\end{equation}
The familiar recurrence relation
for FD and BE functions is valid for the CS functions in general:
\begin{equation}
  \label{eq:103}
  z\frac{\partial}{\partial z}G_{n+1}(z,g)=G_{n}(z,g).
\end{equation}
We also use (\ref{eq:103}) to extend the definition (\ref{eq:53}) to $n\leq0$.

%%%%%%%%%%%%%%%%%%%%%%%%%%%%%%%%%%%%%%%%%%%%%%%
%
\subsection{Equation of state}\label{sec:eos}  
%
%%%%%%%%%%%%%%%%%%%%%%%%%%%%%%%%%%%%%%%%%%%%%%%
The functional relationship between $pV/ \mathcal{N}k_BT$
and fugacity $z$, 
\begin{equation}
  \label{eq:85}
  \frac{pV}{\mathcal{N}k_BT}=
  \frac{G_{\mathcal{D}/2+1}(z,g)}{G_{\mathcal{D}/2}(z,g)},
\end{equation}
\begin{widetext}
%%%%%%%%%%%%%%%%%%%%%%%%%%%%%%%%%%%%%%%%%%%%
\begin{figure}[!h]
  \centering
  \begin{minipage}[c]{0.5\textwidth}
     \centering \includegraphics[width=80mm]{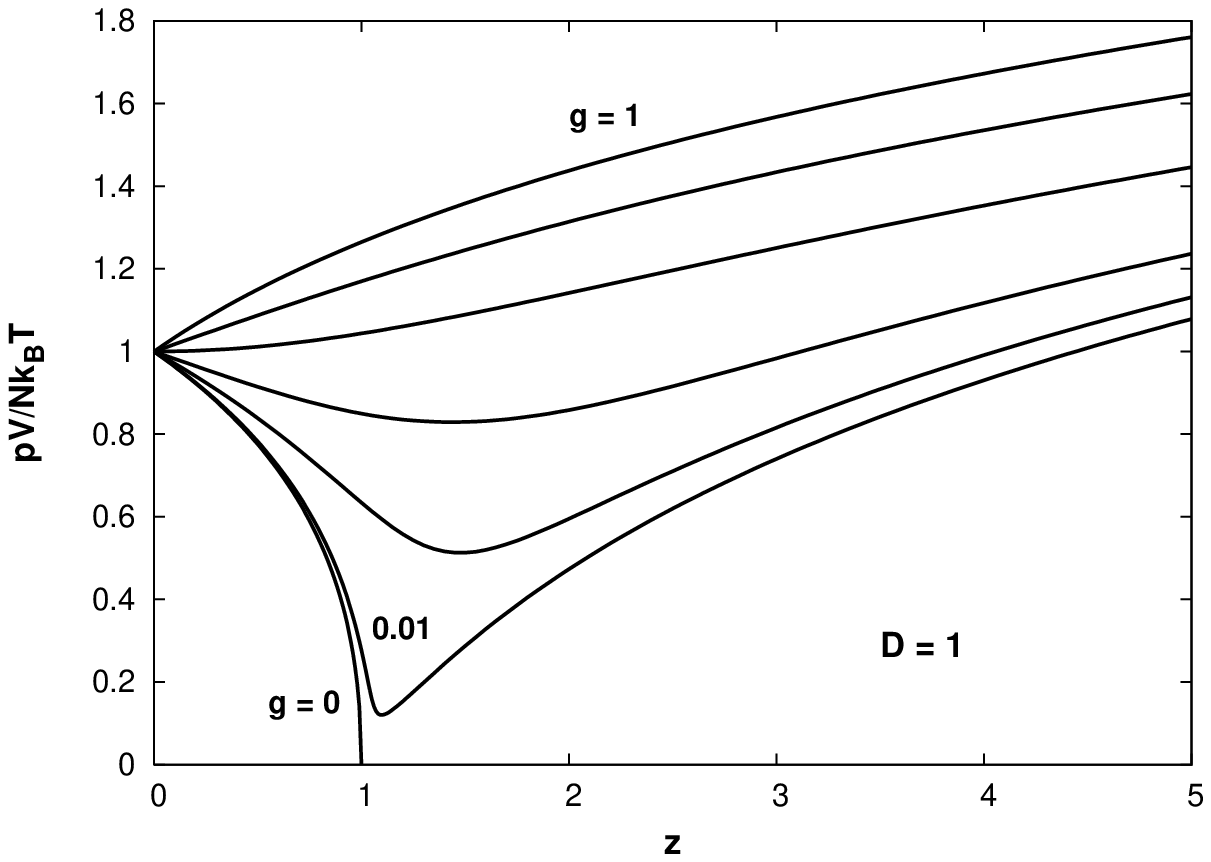}
  \end{minipage}%
  \begin{minipage}[c]{0.5\textwidth}
     \centering \includegraphics[width=80mm]{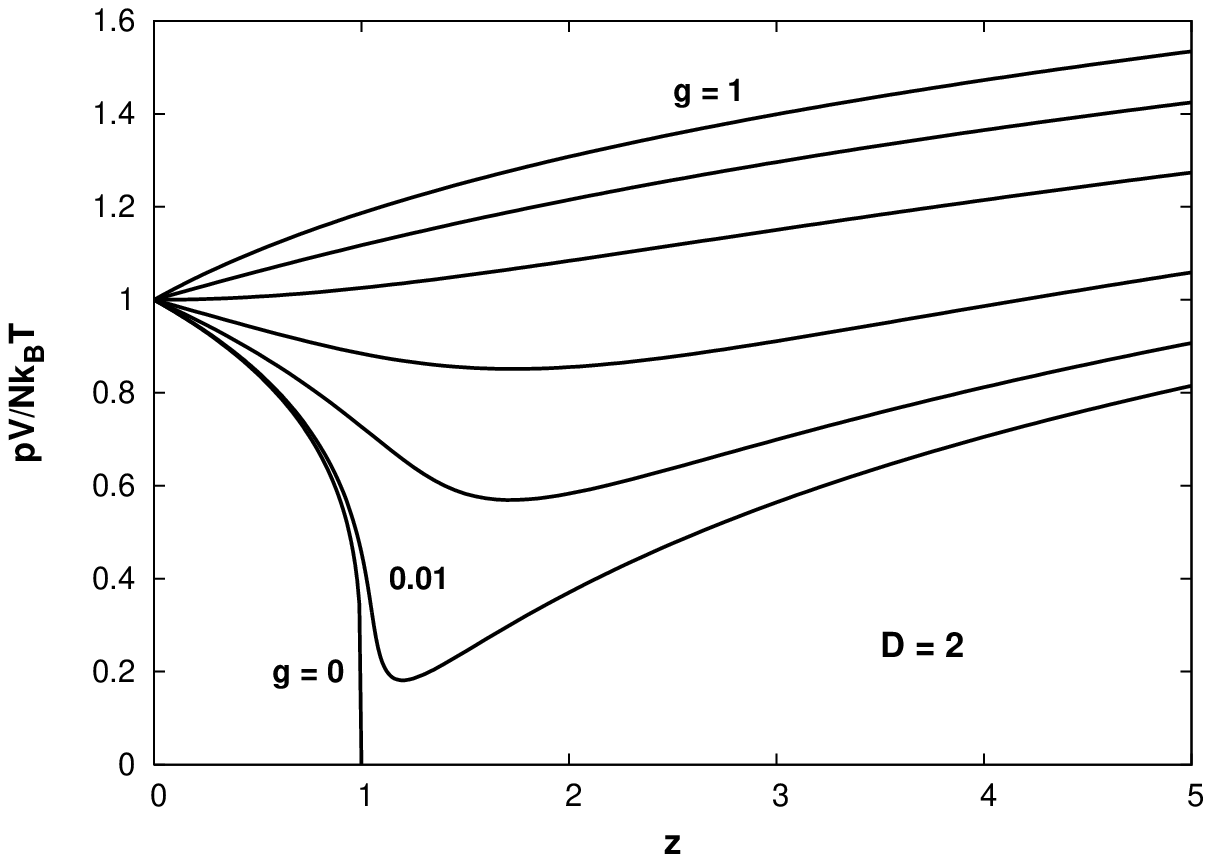}
  \end{minipage}\\
   \begin{minipage}[c]{.5\textwidth}
     \centering \includegraphics[width=80mm]{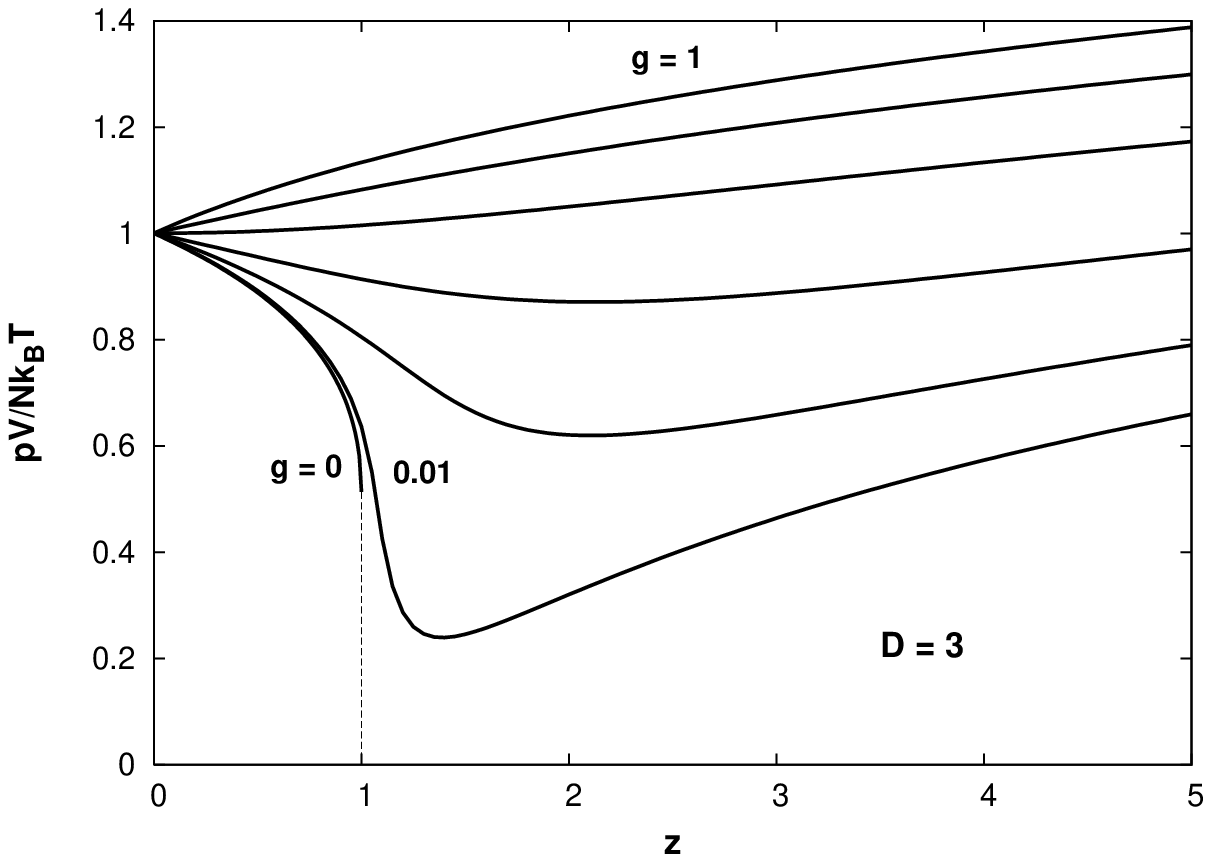}
  \end{minipage}%
  \begin{minipage}[c]{.5\textwidth}
     \centering \includegraphics[width=80mm]{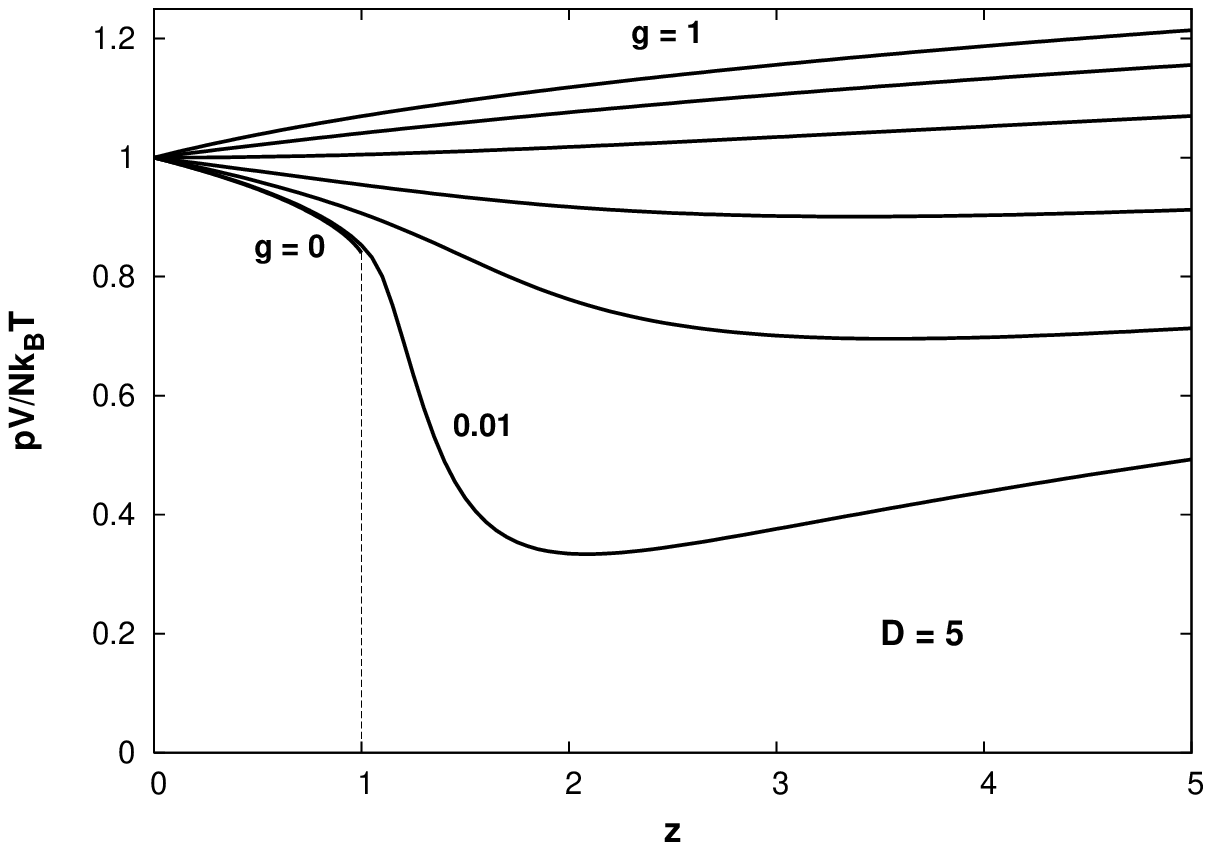}
  \end{minipage}
  \caption{Equation of state (\ref{eq:85}) in
    $\mathcal{D}=1,2,3,5$ for $g=0$, $0.01$, $0.1$, $0.25$, $0.5$, $0.75$, $1$
    (from bottom up). Note the different vertical scales.}\label{fig:eqosD123}
\end{figure}
%%%%%%%%%%%%%%%%%%%%%%%%%%%%%%%%%%%%%%%%%%%%%
\end{widetext}
as inferred from (\ref{eq:50}) and (\ref{eq:52}), highlights the deviations from
Maxwell-Boltzmann (MB) behavior. Numerical results of $pV/ \mathcal{N}k_BT$
versus $z$ are presented in Fig.~\ref{fig:eqosD123}. Increasing $z$ means, for
example, decreasing $T$ at fixed $\mathcal{N}/V$ or increasing $\mathcal{N}/V$
at fixed $T$. Any downward (upward) deviation from $pV/ \mathcal{N}k_BT=1$ is
suggestive of a boson-like (fermion-like) feature.

The curves for $g>\frac{1}{2}$ exhibit a fermion-like monotonic increase over
the entire range of $z$. The curves for $0<g<\frac{1}{2}$ start out with
(boson-like) negative slope. They reach a smooth minimum and then grow without
bound in fermion-like manner.  The case $g=0$ is exceptional. Here the curve
ends in a cusp singularity at $z=1$. In $\mathcal{D}=1$ and $\mathcal{D}=2$ the
pressure is zero at the singularity. The endpoint of the curve occurs at $T=0$
or $\mathcal{N}/V=\infty$. In $\mathcal{D}> 2$ the pressure is nonzero at the
singularity, and the endpoint of the curve signals the onset of a Bose-Einstein
condensate (BEC). For fixed $\mathcal{N}/V$ this happens at $T>0$. The
condensation proceeds along the dashed line.

The physics underlying the crossover between boson-like and fermion-like
behavior may be interpreted by attributing to the statistical interaction of the
generalized CS model a long-range attractive part and a shorter-range repulsive
part. The repulsive core is present for all $g>0$.  The attractive part is only
perceptible for $g<\frac{1}{2}$. The former prevents the collapse of the system
into a BEC at $g>0$. The latter causes a negative interaction pressure for
$g<\frac{1}{2}$ (a reduction of $p$ relative to the kinetic pressure of the MB
gas) if the average interparticle distance in units of the thermal wavelength is
sufficiently large.

The interplay between the two parts of the statistical interaction upon
variation of the parameters $g$ and $\mathcal{D}$ produces a host of
interesting thermodynamic effects. Their appearance in isochores, isotherms,
isobars, response functions, and in the velocity of sound will be discussed in
the following.  For this purpose we introduce reference values for temperature
$T$, reduced volume $v\doteq V/ \mathcal{N}$, and pressure $p$, based on the
equation of state, $pv=k_BT$ of the MB gas and the criterion that the average
reference volume per particle is a hypercubic box with sides equal to the
thermal wavelength (\ref{eq:10}). For isochoric, isothermal, and isobaric processes
we thus use
%\vspace*{-5mm}
\begin{align*}
%  \label{eq:108}
  k_BT_v = \frac{4\pi}{v^{2/ \mathcal{D}}},\quad p_v = \frac{4\pi}{v^{2/
        \mathcal{D}+1}}\qquad (v = \mathrm{const.}) 
%\end{align}
%\begin{align}
\\  % \label{eq:109}
  v_T = \left(\frac{4\pi}{k_BT}\right)^{\mathcal{D}/2},\quad
  p_T = \frac{(k_BT)^{\mathcal{D}/2+1}}{(4\pi)^{\mathcal{D}/2}}\qquad 
(T = \mathrm{const.})
%\end{align}
%\begin{align}
 \\  % \label{eq:110}
  k_BT_p = (4\pi)^{\frac{\mathcal{D}}{\mathcal{D}+2}}p^{\frac{2}{\mathcal{D}+2}},\quad  
v_p =  (4\pi/p)^{\frac{\mathcal{D}}{\mathcal{D}+2}}\qquad 
(p = \mathrm{const.})
\end{align*}
%\vspace{-5mm}
These reference values are well-behaved in the boson limit, which is an
advantage for our comparative plots. For some purposes (limited to $g>0$) an
alternative choice, based on the chemical potential at $T=0$, will be more convenient.
%%%%%%%%%%%%%%%%%%%%%%%%%%%%%%%%%%%%%%%%%%%%%%%
%
\subsection{Isochores}\label{sec:isoc}  
\begin{widetext}
%%%%%%%%%%%%%%%%%%%%%%%%%%%%%%%%%%%%%%%%%%%%
\begin{figure}[ht!]
  \centering
  \begin{minipage}[c]{0.5\textwidth}
     \centering \includegraphics[width=85mm]{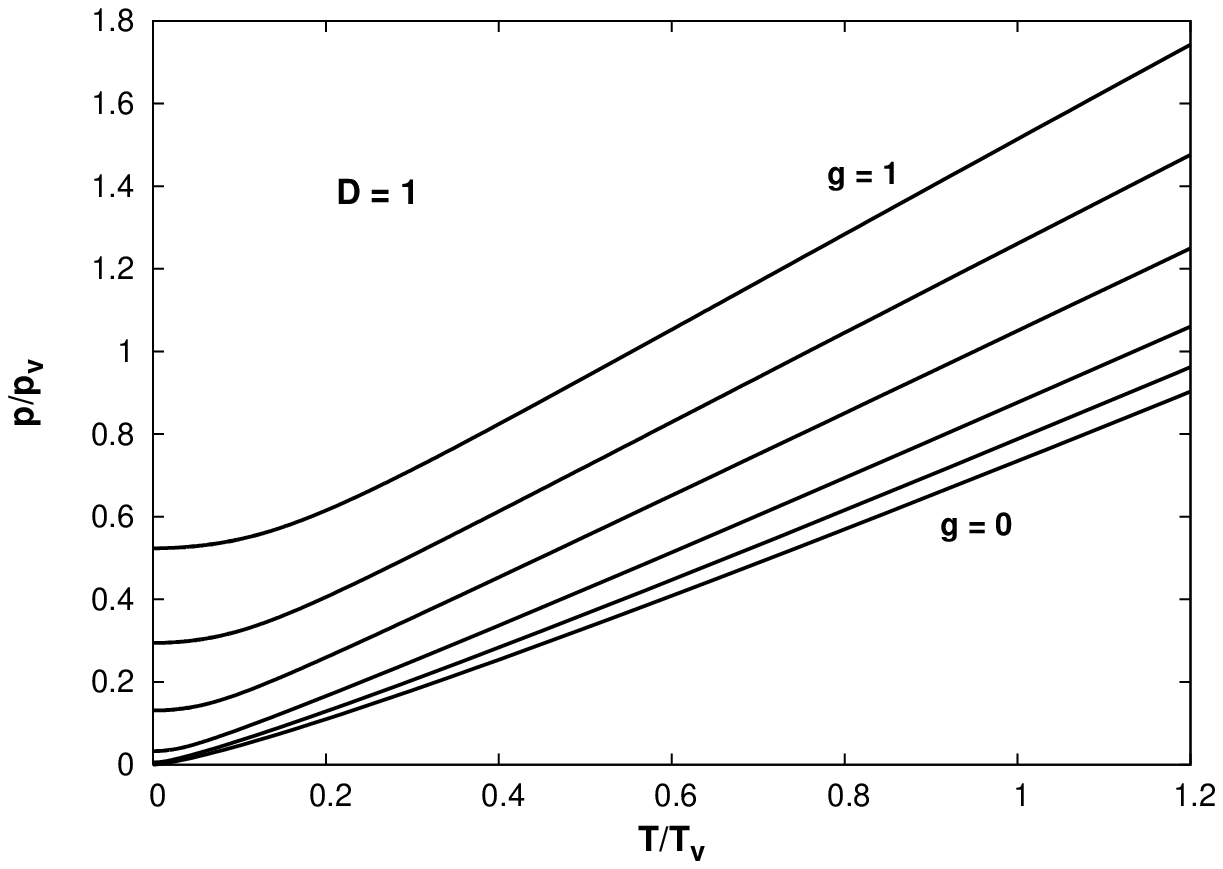}
  \end{minipage}%
  \begin{minipage}[c]{0.5\textwidth}
     \centering \includegraphics[width=85mm]{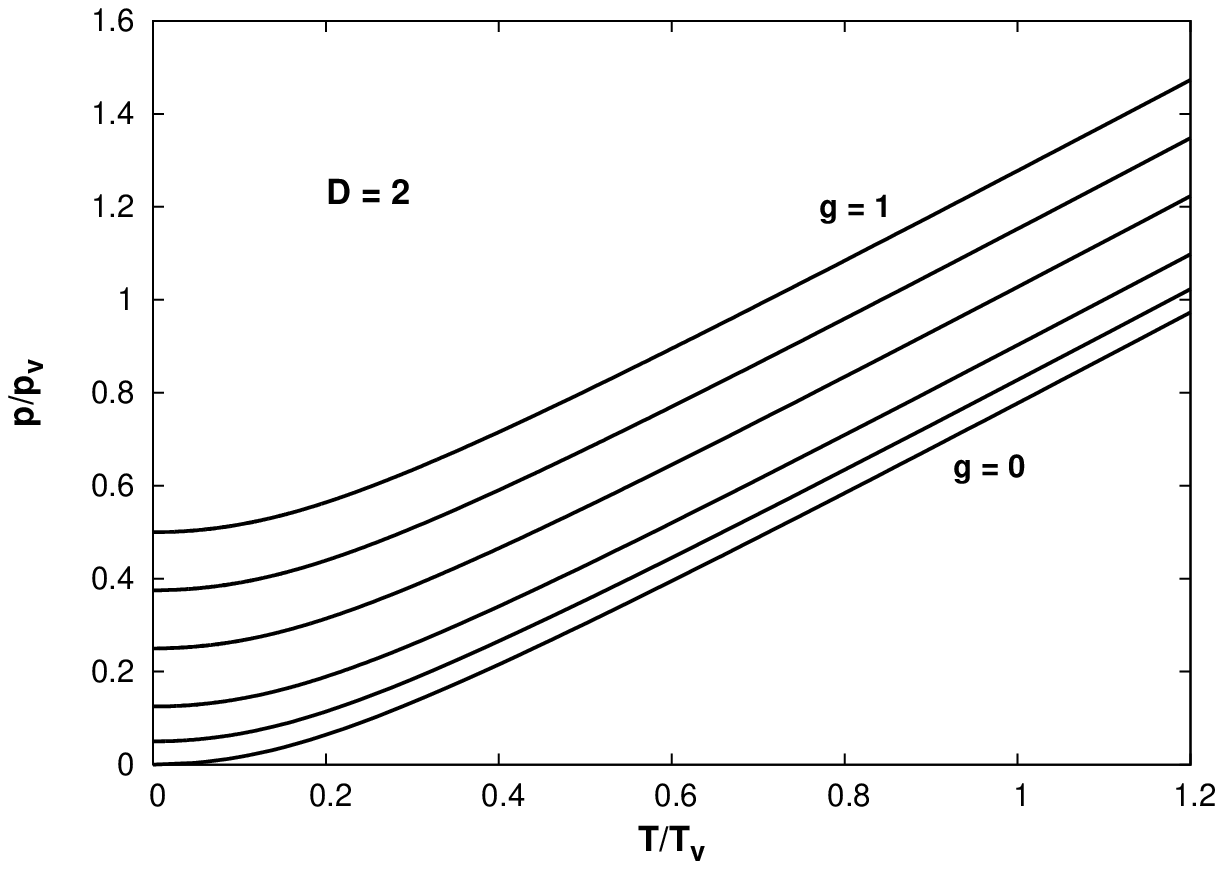}
  \end{minipage}\\
   \begin{minipage}[c]{.5\textwidth}
     \centering \includegraphics[width=85mm]{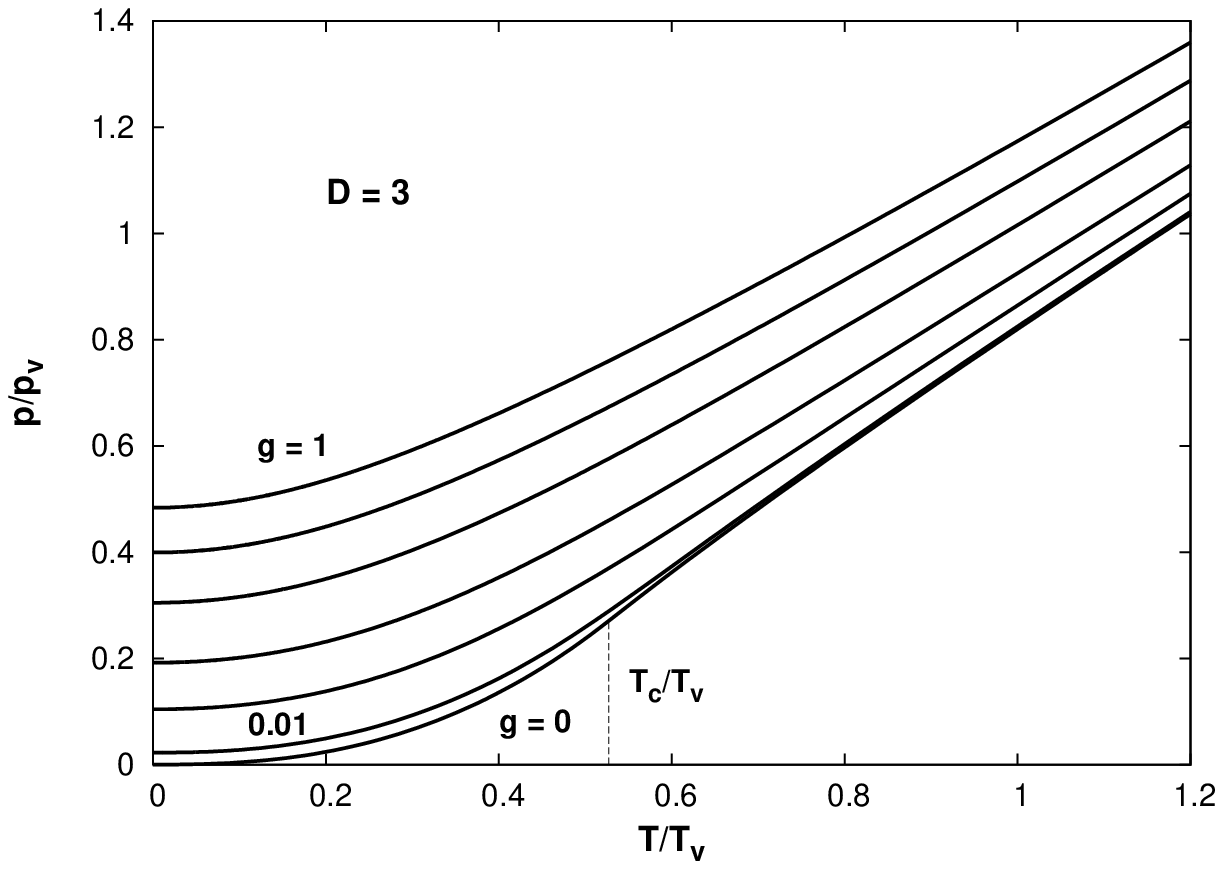}
  \end{minipage}%
  \begin{minipage}[c]{.5\textwidth}
     \centering \includegraphics[width=85mm]{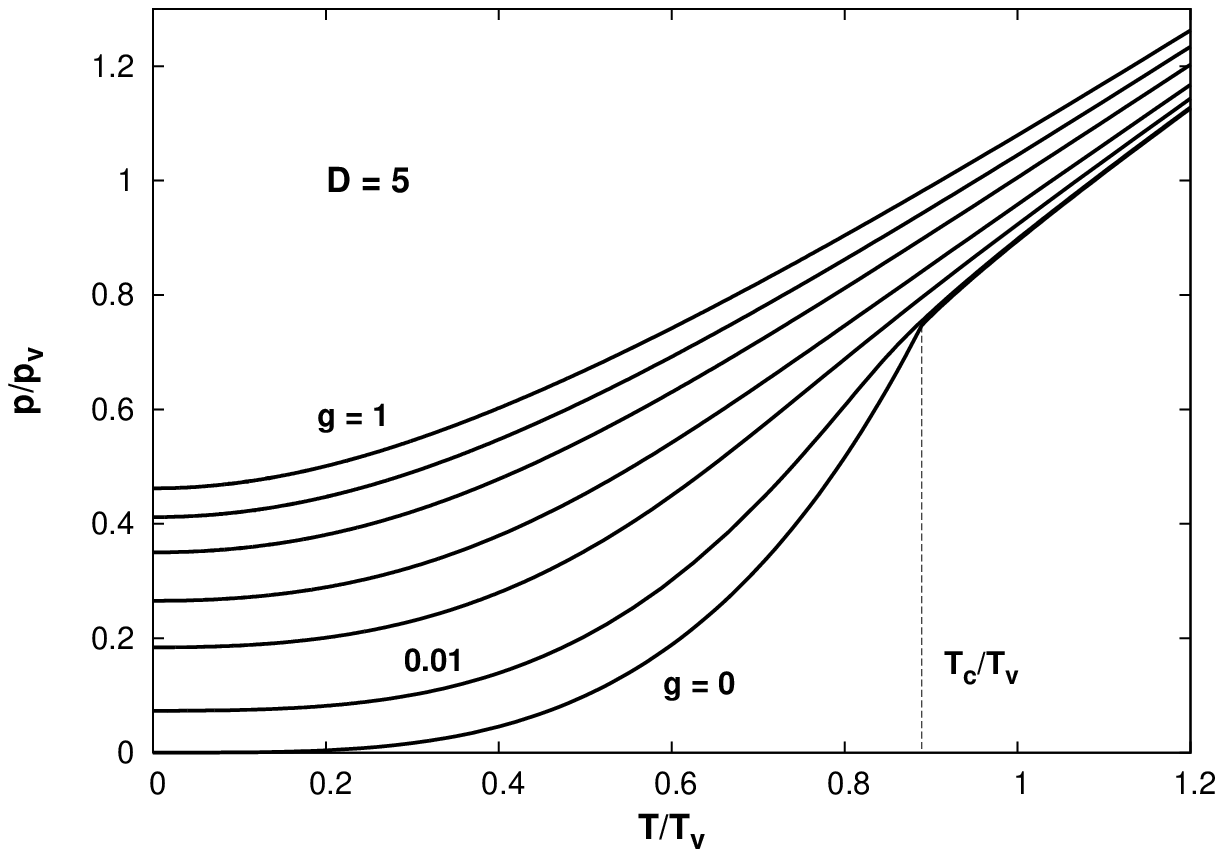}
  \end{minipage}
  \caption{Isochores in $\mathcal{D} =1,2,3,5$ for $g=0$, $0.01$, $0.1$,
    $0.25$, $0.5$, $0.75$, $1$ (from bottom up). In $\mathcal{D}=1,2$ the
    curves for $g=0$, $0.01$ are unresolved. Note the different vertical scales.}
  \label{fig:gencsppv}
\end{figure}
%%%%%%%%%%%%%%%%%%%%%%%%%%%%%%%%%%%%%%%%%%%%%
\end{widetext}

%
%%%%%%%%%%%%%%%%%%%%%%%%%%%%%%%%%%%%%%%%%%%%%%%
The dependence of $p$ on $T$ at constant $v$ can be extracted from
(\ref{eq:50}) and (\ref{eq:52}):
\begin{subequations}
    \label{eq:111}
    \begin{align}
      \frac{p}{p_v} & = \frac{G_{\mathcal{D}/2+1}(z,g)}{[G_{\mathcal{D}/2}(z,g)]^{1+2/
          \mathcal{D}}},
      \\
      \frac{T}{T_v} & = \frac{1}{[G_{\mathcal{D}/2}(z,g)]^{2/ \mathcal{D}}} 
\end{align}
\end{subequations}
with no restriction on $z$ for $0<g\leq1$.  In Fig.~\ref{fig:gencsppv} we show
isochores for various $g$ and $\mathcal{D}$. Each curve represents a universal
isochore, valid for arbitrary values of $v$.
The shape of the curves at high $T$ reflects the emerging MB behavior. The
leading correction term in the high-$T$ expansion of (\ref{eq:111}),
 \begin{equation}
    \label{eq:115}
    \frac{p}{p_v} \stackrel{T\to\infty}{\leadsto} \frac{T}{T_v}\left[1- \frac{1/2-g}{2^{\mathcal{D}/2}}
      \left(\frac{T_v}{T}\right)^{\mathcal{D}/2}\right],
\end{equation}
describes bundles of isochores whose vertical separations increase with
increasing $T/T_v$ if $\mathcal{D}<2$, stay constant if $\mathcal{D}=2$, and
decrease if $\mathcal{D}>2$. The deviation from the MB isochore, $p/p_v=T/T_v$,
is negative for $g<\frac{1}{2}$ and positive for $g>\frac{1}{2}$, decreasing in
magnitude as $\mathcal{D}$ increases. For semions the deviation is of higher
order in $T_v/T$.
  
The intercept of the isochore at $T=0$ as extracted from the leading term in the
asymptotic expansion of (\ref{eq:111}) is
\begin{subequations}
    \label{eq:116}  
\begin{align} 
  \frac{p_0}{p_v} & = \frac{\bar{T}_v}{T_v}\frac{1}{\mathcal{D}/2+1},
  \\
  \frac{\bar{T}_v}{T_v} & = \left[g\Gamma(\mathcal{D}/2+1)\right]^{2/ \mathcal{D}},
\end{align}
\end{subequations}
where $\mu_0=k_B\bar{T}_v$ is the chemical potential at $T=0$.  If we replot the
data for $g>0$ in Fig. \ref{fig:gencsppv} as $p/ \bar{p}_v$ versus $T/
\bar{T}_v$ with $\bar{p}_v=p_v(T/ \bar{T}_v)$ we obtain the MB isochore in the
limit $\mathcal{D}\to\infty$.

In the boson case, Eq.~(\ref{eq:111}) is still valid for $z<1$. This covers the
full $T$-range in $\mathcal{D}\leq2$. In $\mathcal{D}>2$ in the limit $z\to1$
occurs at
\begin{subequations}
  \label{eq:112}
  \begin{align}
    \frac{p_c}{p_v} & =
    \frac{\zeta(\mathcal{D}/2+1)}{[\zeta(\mathcal{D}/2)]^{2/\mathcal{D}+1}},
    \\ 
    \frac{T_c}{T_v} & = \frac{1}{[\zeta(\mathcal{D}/2)]^{2/ \mathcal{D}}},
\end{align}
\end{subequations}
where $G_n(1,0)=\zeta(n)$ is the Riemann zeta function. The values $p_c,T_c$ mark
the onset of a BEC. Note that $T_c$ begins to rise from zero at
$\mathcal{D}=2$ and approaches the value $T_v$ as $\mathcal{D}\to\infty$. The
bosonic isochore in the coexistence region is
\begin{equation}
  \label{eq:113}
  \frac{p}{p_v}=\left(\frac{T}{T_v}\right)^{\mathcal{D}/2+1}\zeta(\mathcal{D}/2+1)
  \qquad (T<T_c).
\end{equation}
The singularity at $T_c>0$ becomes stronger as $\mathcal{D}$ increases. For
$2<\mathcal{D}\leq4$, there is a discontinuity in curvature. It turns into a
discontinuity in slope for $\mathcal{D}>4$ and becomes a discontinuity in the
function itself for $\mathcal{D}=\infty$.

%%%%%%%%%%%%%%%%%%%%%%%%%%%%%%%%%%%%%%%%%%%%%%%
%
\subsection{Isotherms}\label{sec:isot}  
\begin{widetext}

%%%%%%%%%%%%%%%%%%%%%%%%%%%%%%%%%%%%%%%%%%%%
\begin{figure}[ht!]
  \centering
  \begin{minipage}[c]{0.5\textwidth}
     \centering \includegraphics[width=85mm]{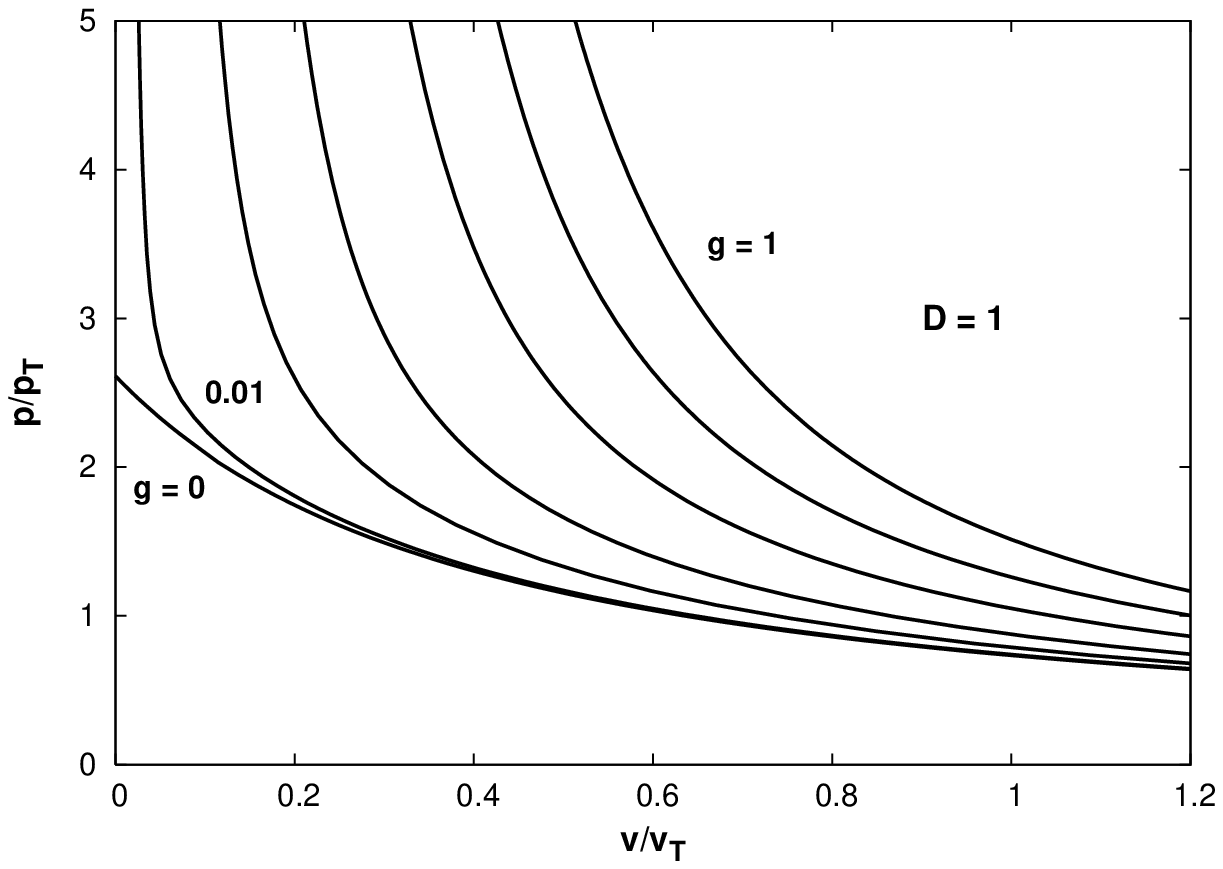}
  \end{minipage}%
  \begin{minipage}[c]{0.5\textwidth}
     \centering \includegraphics[width=85mm]{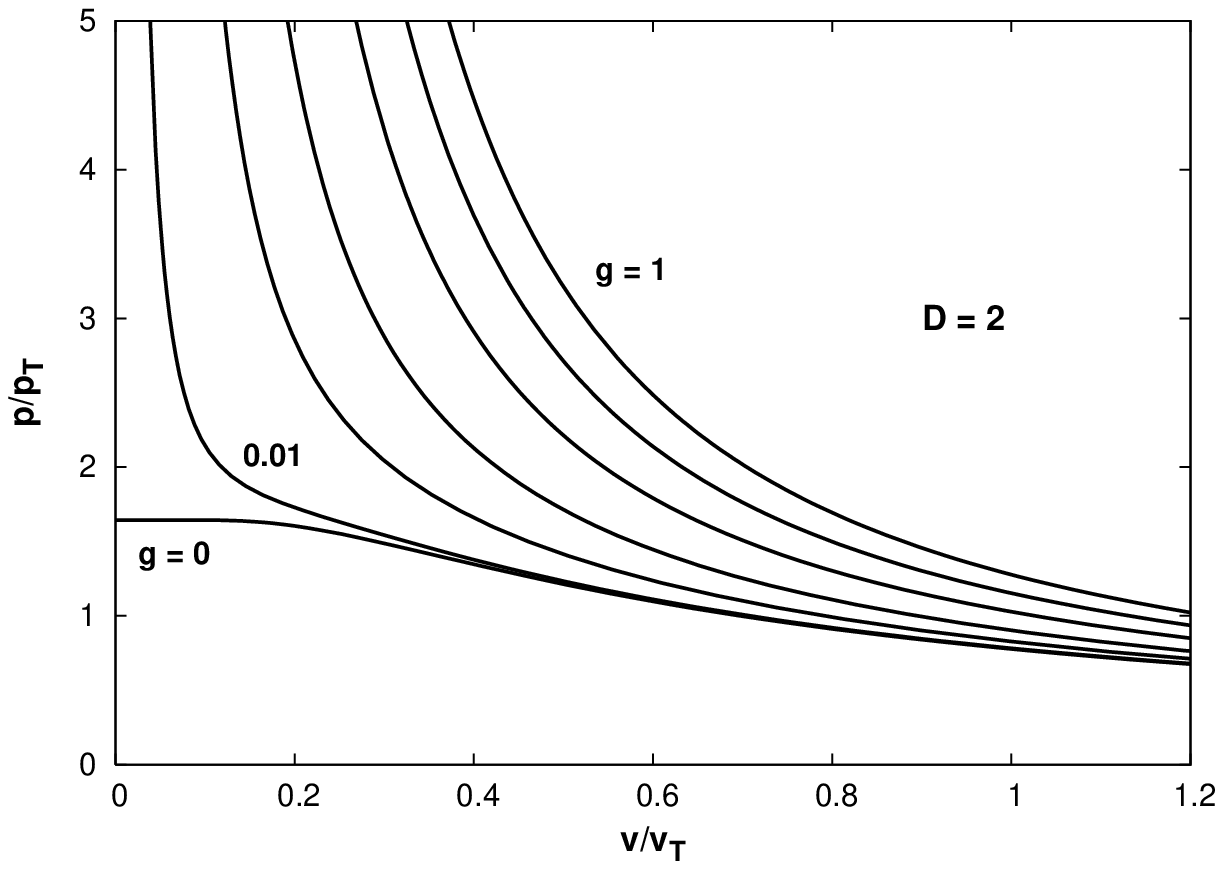}
  \end{minipage}\\
  \begin{minipage}[c]{0.5\textwidth}
     \centering \includegraphics[width=85mm]{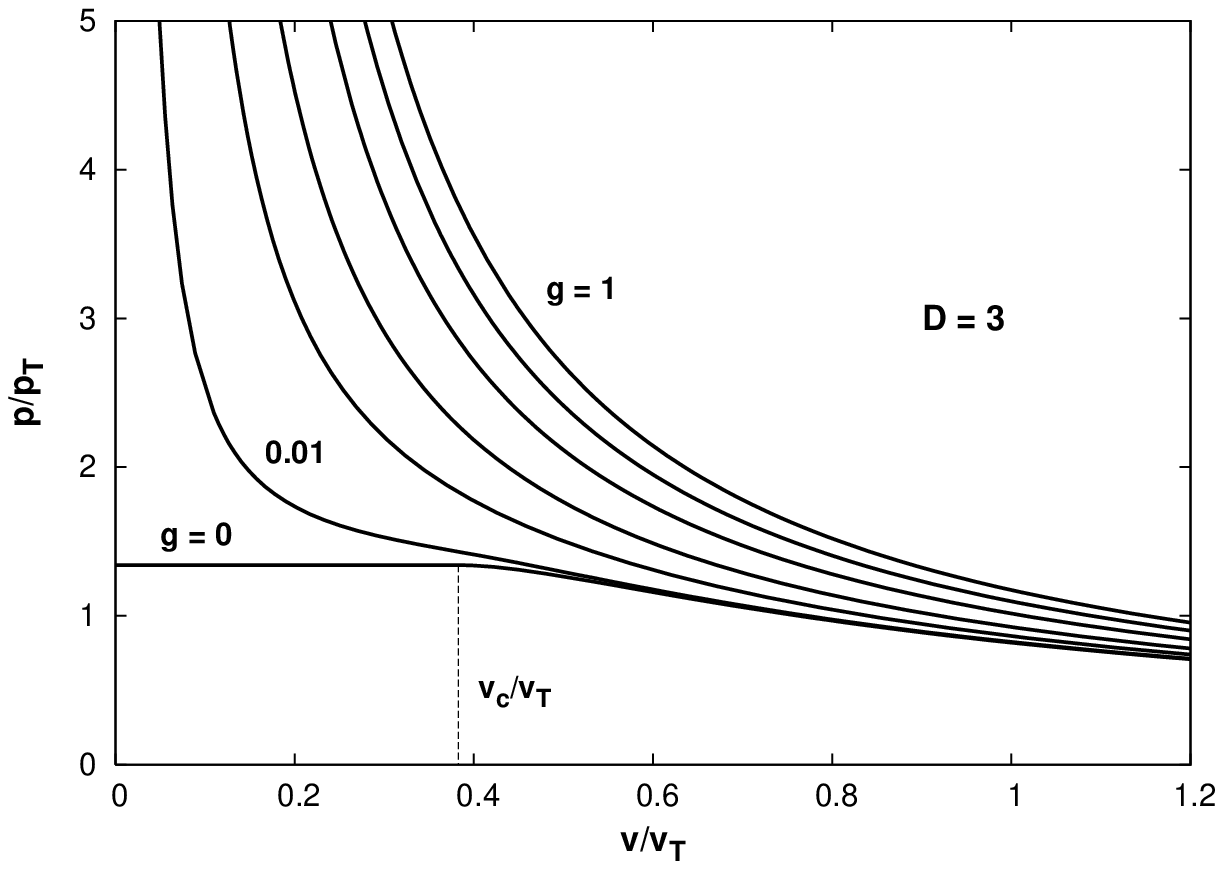}
  \end{minipage}%
  \begin{minipage}[c]{0.5\textwidth}
    \centering \includegraphics[width=85mm]{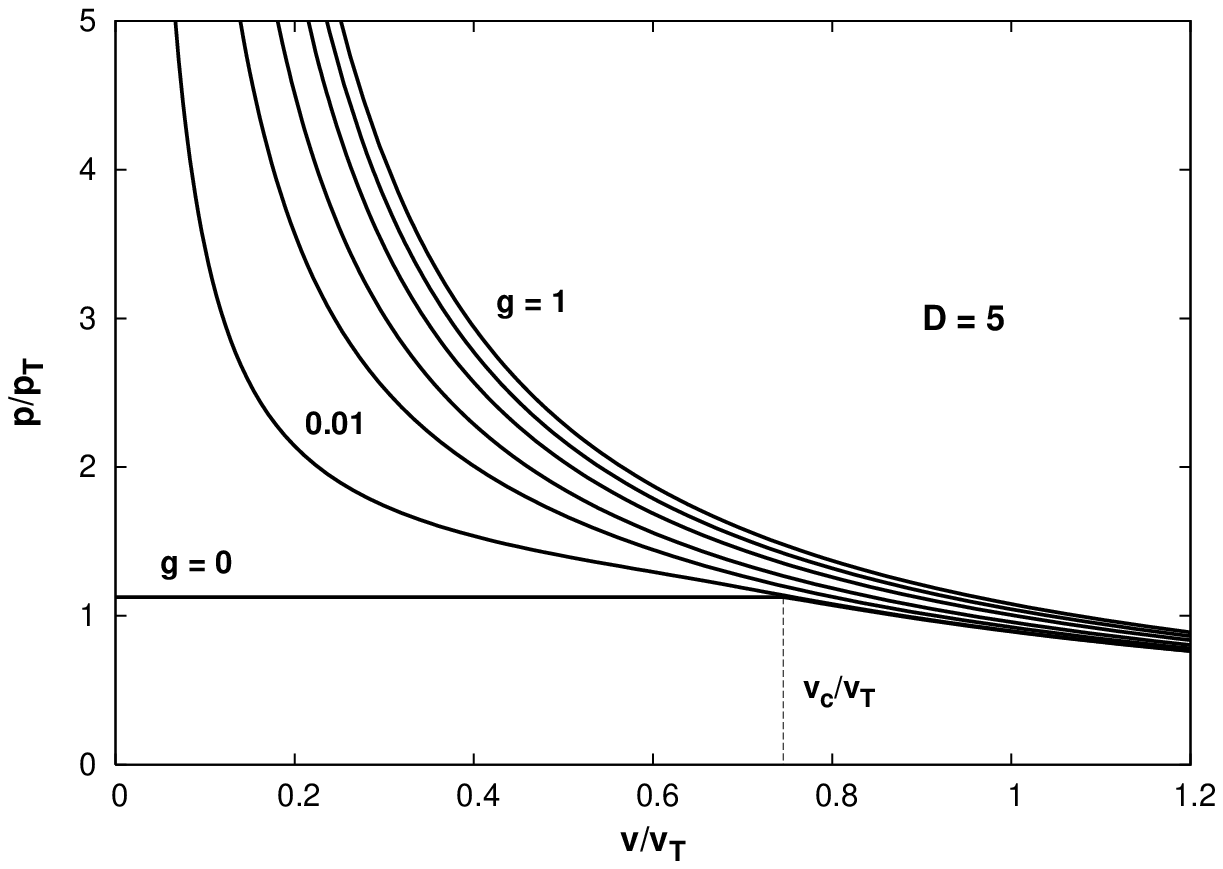}
  \end{minipage}
  \caption{Isotherms of the generalized CS model in dimensions $\mathcal{D} =
    1,2,3,5$ for $g=0$, $0.01$, $0.1$, $0.25$, $0.5$, $0.75$, $1$ (from bottom
    up).}
  \label{fig:gencsisoth}
\end{figure}
%%%%%%%%%%%%%%%%%%%%%%%%%%%%%%%%%%%%%%%%%%%%%
\end{widetext}

%
%%%%%%%%%%%%%%%%%%%%%%%%%%%%%%%%%%%%%%%%%%%%%%%
The dependence of $p$ on $v$ at constant $T$ in a parametric
representation follows again from (\ref{eq:50}) and (\ref{eq:52}):
\begin{subequations}
  \label{eq:114a}
  \begin{align}
    \frac{p}{p_T} & = G_{\mathcal{D}/2+1}(z,g),
    \\ 
    \frac{v}{v_T} & = \frac{1}{G_{\mathcal{D}/2}(z,g)}
  \end{align}
\end{subequations}
applicable for $0<g\leq1$ with unrestricted $z$ and for $g=0$ with $z\leq 1$.  In
Fig.~\ref{fig:gencsisoth} we show isotherms for various $g$ and $\mathcal{D}$.
The convergence of the curves in the low-density regime reflects the emerging
MB behavior. The leading correction to Boyle's law,
  \begin{equation}
    \label{eq:123}
    \frac{p}{p_T} \stackrel{v\to\infty}{\leadsto} 
    \frac{v_T}{v}\left[1- \frac{1/2-g}{2^{\mathcal{D}/2}}
      \left(\frac{v_T}{v}\right)\right],
  \end{equation}
  is negative for $g<\frac{1}{2}$ and positive for $g>\frac{1}{2}$. It weakens
  in magnitude as $\mathcal{D}$ increases. The divergence of the isochores for
  $g>0$ in the high-density limit is of the form
\begin{equation}
  \label{eq:125}
  \left(\frac{p}{p_T}\right)\left(\frac{v}{v_T}\right)^{(\mathcal{D}+2)/ \mathcal{D}} 
=g^{2/ \mathcal{D}} 
\frac{[\Gamma(\mathcal{D}/2+1)]^{2/ \mathcal{D}}}{\mathcal{D}/2+1},
\end{equation}
consistent with MB behavior for $g>0$ in the limit $\mathcal{D}\to\infty$,
if we switch to alternative reference values based on the chemical potential.
 
For bosons the pressure $p/p_T$ remains finite as $v/v_T\to0$.  
In $\mathcal{D}\leq2$ the limit $z\to1$ implies 
\begin{equation}
  \label{eq:126z}
  \frac{v}{v_T} \to 0,\quad \frac{p}{p_T} \to \zeta\left(\frac{\mathcal{D}}{2}+1\right),
\end{equation}
whereas in $\mathcal{D}>2$ at $z=1$ we have
\begin{equation}
  \label{eq:115a}
  \frac{v_c}{v_T}= \frac{1}{\zeta(\mathcal{D}/2)},
  \quad  
  \frac{p_c}{p_T}= \zeta(\mathcal{D}/2+1).
\end{equation}
The intercept of the bosonic isotherm decreases with increasing $\mathcal{D}$
and approaches unity for $\mathcal{D}\to\infty$.  In $\mathcal{D}\leq2$ the
bosonic isotherm is a smooth curve. Its slope at $v/v_T=0$ is negative in
$\mathcal{D}<2$ and zero in $\mathcal{D}=2$. In $\mathcal{D}>2$ the pressure is
constant at $p_c$ for $0\leq v\leq v_c$ along the isotherm. In the limit
$\mathcal{D}\to\infty$ we have the MB isotherm, $p/p_T=v_T/v$ at $v>v_T$ joined by a
the constant $p/p_T=1$ at $v/v_T<1$.

%%%%%%%%%%%%%%%%%%%%%%%%%%%%%%%%%%%%%%%%%%%%%%%
%
\subsection{Isobars}\label{sec:isob}  
\begin{widetext}
%%%%%%%%%%%%%%%%%%%%%%%%%%%%%%%%%%%%%%%%%%%%
\begin{figure}[t!]
  \centering
  \begin{minipage}[c]{0.5\textwidth}
     \centering \includegraphics[width=85mm]{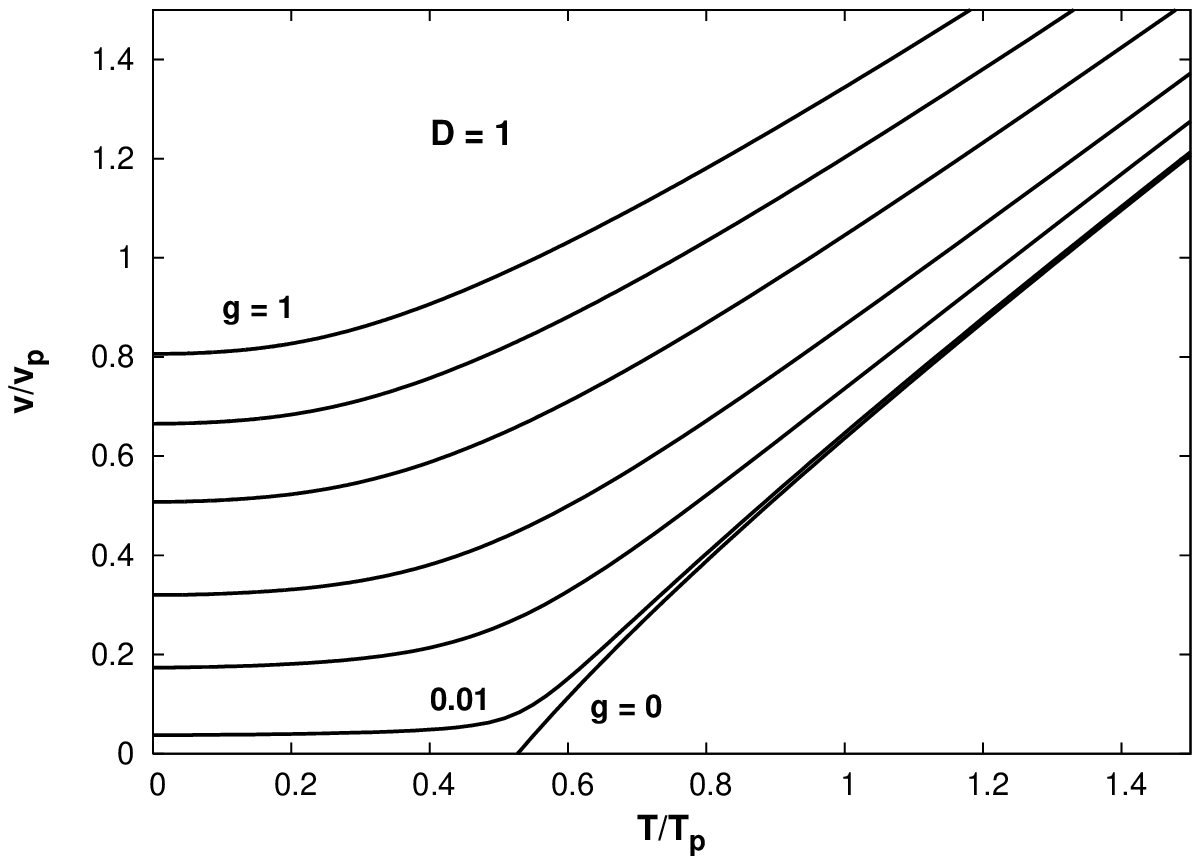}
  \end{minipage}%
  \begin{minipage}[c]{0.5\textwidth}
     \centering \includegraphics[width=85mm]{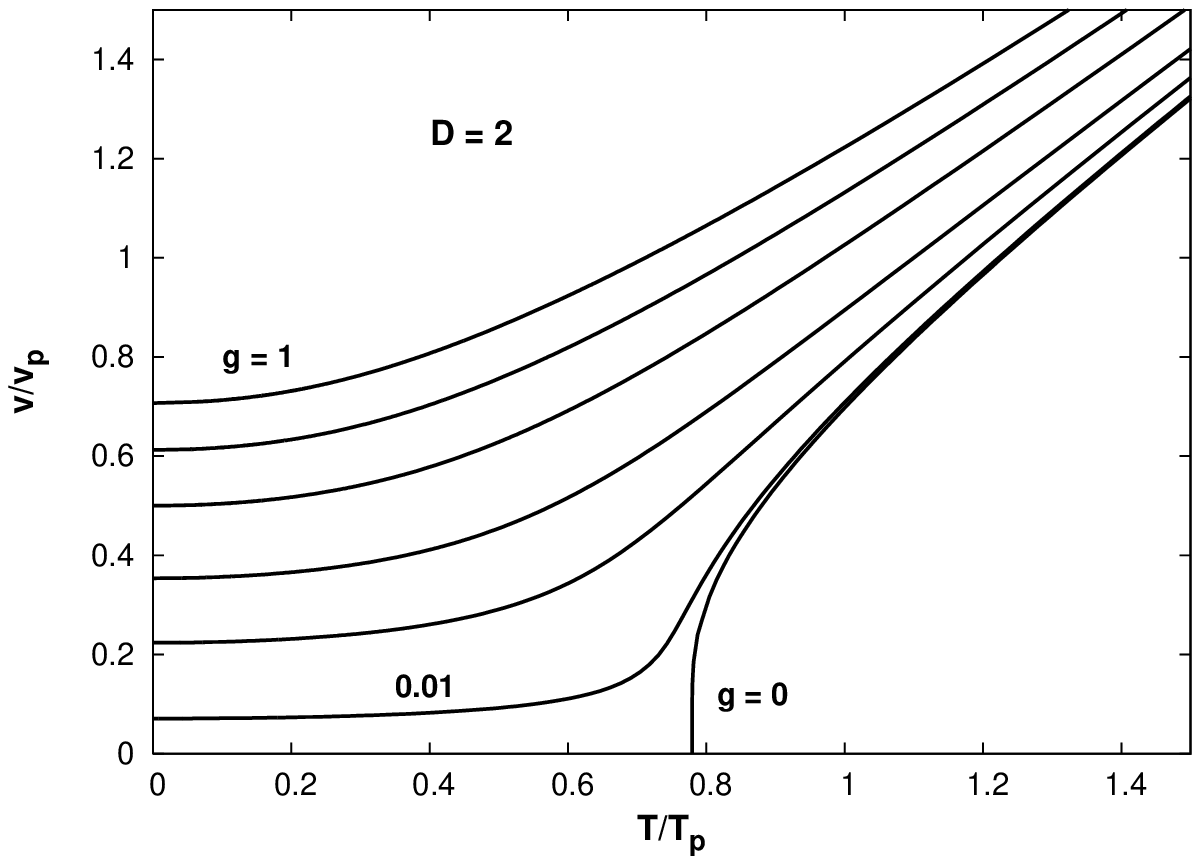}
  \end{minipage}\\
   \begin{minipage}[c]{.5\textwidth}
     \centering \includegraphics[width=85mm]{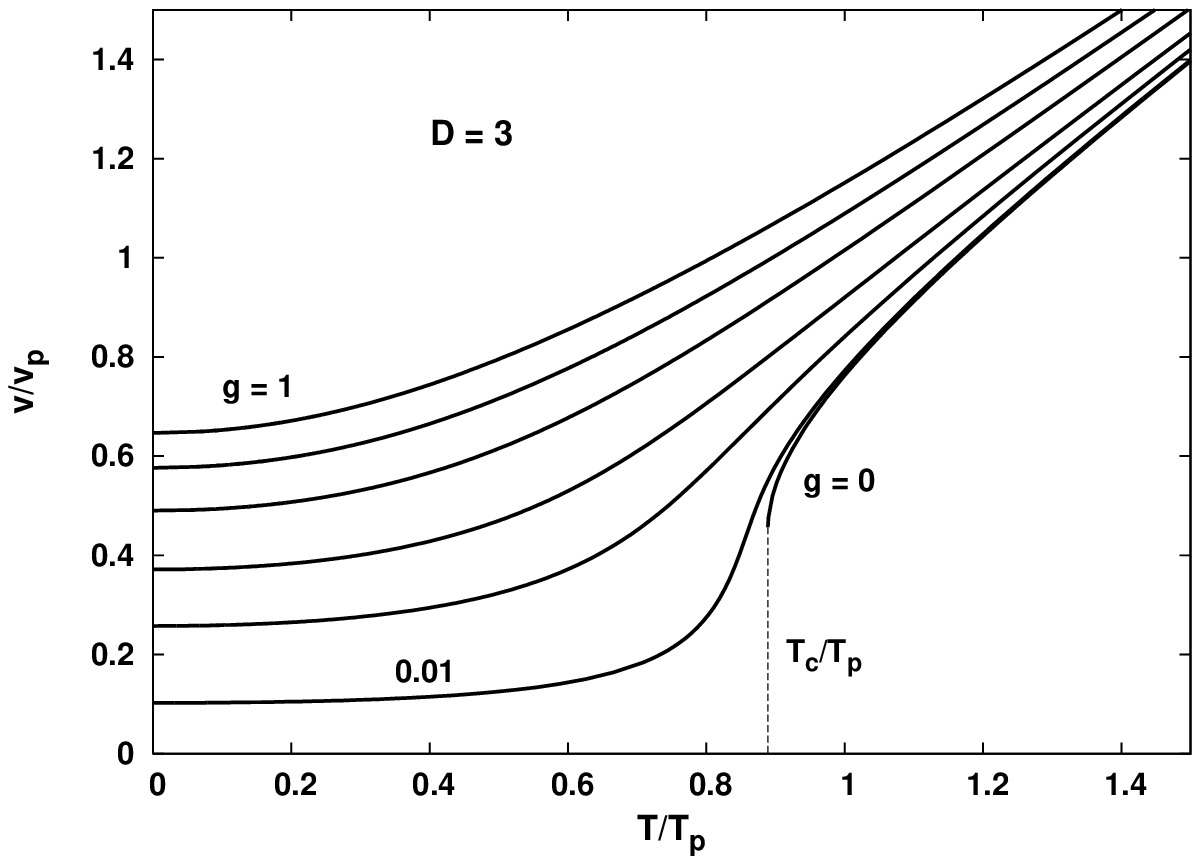}
  \end{minipage}%
  \begin{minipage}[c]{.5\textwidth}
     \centering \includegraphics[width=85mm]{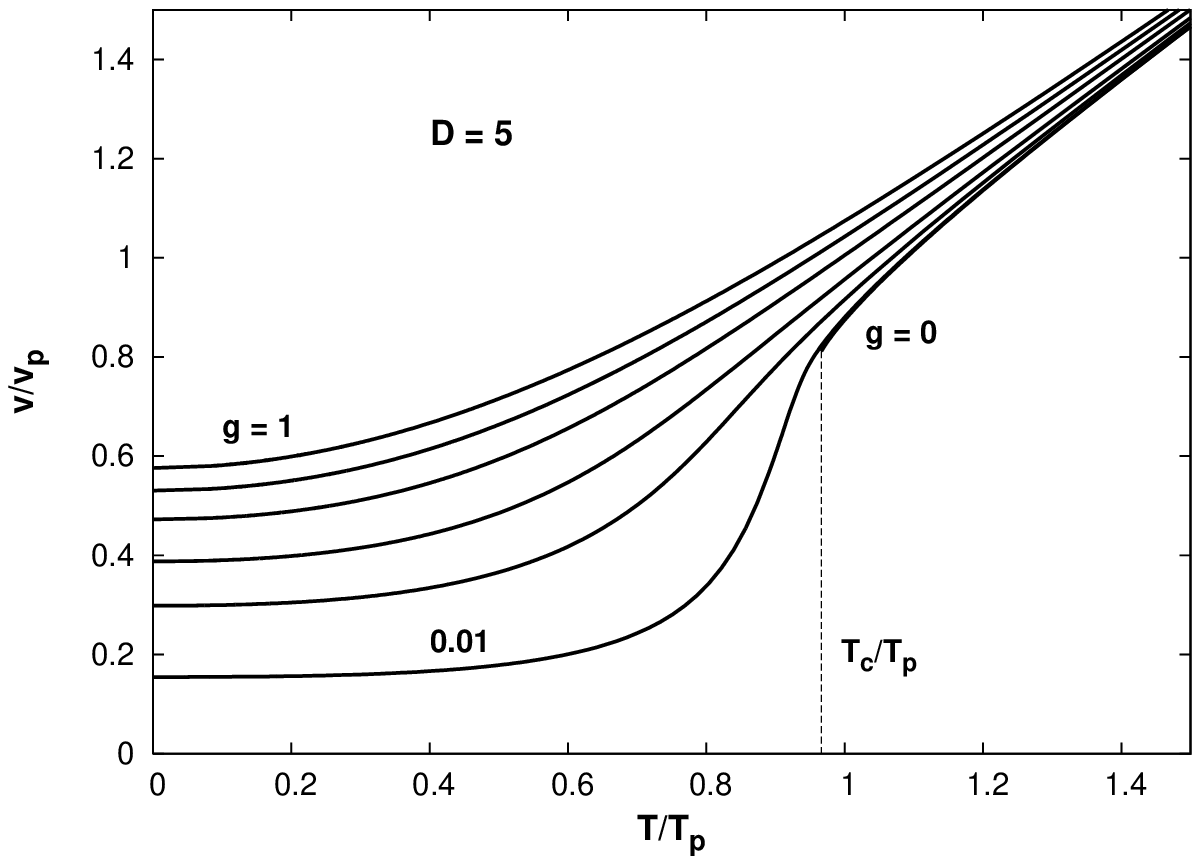}
  \end{minipage}
  \caption{Isobars in $\mathcal{D} = 1,2,3,5$ for $g=0$, $0.01$, $0.1$, $0.25$,
    $0.5$, $0.75$, $1$ (from bottom up).}
  \label{fig:gencsisob}
\end{figure}
%%%%%%%%%%%%%%%%%%%%%%%%%%%%%%%%%%%%%%%%%%%%%
\end{widetext}

%
%%%%%%%%%%%%%%%%%%%%%%%%%%%%%%%%%%%%%%%%%%%%%%%
The dependence of $v$ on $T$ at constant $p$ is to be calculated from
\begin{subequations}
  \label{eq:27}
  \begin{align}
    \label{eq:27a}
    \frac{v}{v_p} & =
    \frac{[G_{\mathcal{D}/2+1}(z,g)]^{\mathcal{D}/(\mathcal{D}+2)}}
    {G_{\mathcal{D}/2}(z,g)}, \\     \label{eq:27b}
    \frac{T}{T_p} &=
    \frac{1}{\left[G_{\mathcal{D}/2+1}(z,g)\right]^{2/(\mathcal{D}+2)}}, 
\end{align}
\end{subequations}
applicable for $0<g\leq1$ with unrestricted $z$ and for $g=0$ with $z\leq1$.
In Fig.~\ref{fig:gencsisob} we show isobars for various $g$ and $\mathcal{D}$.
The leading correction to MB behavior at high $T$,
\begin{equation}
    \label{eq:128}
    \frac{v}{v_p} \stackrel{T\to\infty}{\leadsto} 
    \frac{T}{T_p}\left[
      1-\frac{1/2-g}{2^{\mathcal{D}/2}}\left(\frac{T_p}{T}\right)^{\mathcal{D}/2+1}
    \right],
\end{equation}
explains the observation that the vertical separations of the curves in
Fig.~\ref{fig:gencsisob} decrease with increasing $T/T_p$ in all $\mathcal{D}$.
Again the deviation from MB behavior switches sign at $g=\frac{1}{2}$.  The
intercept at $T=0$ of the isobars for $g>0$ is
\begin{equation}
    \label{eq:129}
    \lim_{T\to0}\frac{v}{v_p}=g^{2/(\mathcal{D}+2)}
\frac{\left[\Gamma(\mathcal{D}/2+1)\right]^{2/(\mathcal{D}+2)}}%
{\left(\mathcal{D}/2+1\right)^{\mathcal{D}/(\mathcal{D}+2)}},
\end{equation}
again consistent with MB behavior for $g>0$ in the limit
$\mathcal{D}\to\infty$ provided we use alternative reference values based on the
chemical potential.

The isobaric curves for bosons end in a critical point at
\begin{subequations}
  \label{eq:117}
  \begin{align}
    \frac{v_c}{v_p} & =
    \frac{\left[\zeta(\mathcal{D}/2+1)\right]^{\mathcal{D}/(\mathcal{D}+2)}}%
    {\zeta(\mathcal{D}/2)},
    \\ 
    \frac{T_c}{T_p} & = \frac{1}{\left[\zeta(\mathcal{D}/2+1)\right]^{2/(\mathcal{D}+2)}}.
\end{align}
\end{subequations}
Note that the critical (reduced) volume is nonzero only in $\mathcal{D}>2$
whereas the critical temperature is nonzero in all $\mathcal{D}\geq1$.  At
$T<T_c$ the boson gas is unable to maintain the prescribed pressure. In
$\mathcal{D}\leq2$ the bosonic isobar terminates in a cusp at $v/v_p=0$. The
switch from a pure gas at $T>T_c$ to a pure BEC at $T<T_c$ occurs when
$v/v_p=0$. There is no coexistence region.  In $\mathcal{D}>2$ the bosonic
isobar terminates in a cusp at $T=T_c>0$ and $v=v_c>0$. Here the two phases do
coexist. During condensation $v/v_p$ gradually goes to zero along the vertical
dashed line. With increasing $\mathcal{D}$ both $v_c/v_p$ and $T_c/T_p$
increase toward unity.

%%%%%%%%%%%%%%%%%%%%%%%%%%%%%%%%%%%%%%%%%%%%%%%
%
\subsection{Isochoric heat capacity}\label{sec:heatcap}  
\begin{widetext}
%%%%%%%%%%%%%%%%%%%%%%%%%%%%%%%%%%%%%%%%%%%%
\begin{figure}[t!]
  \centering
  \begin{minipage}[c]{0.5\textwidth}
     \centering \includegraphics[width=85mm]{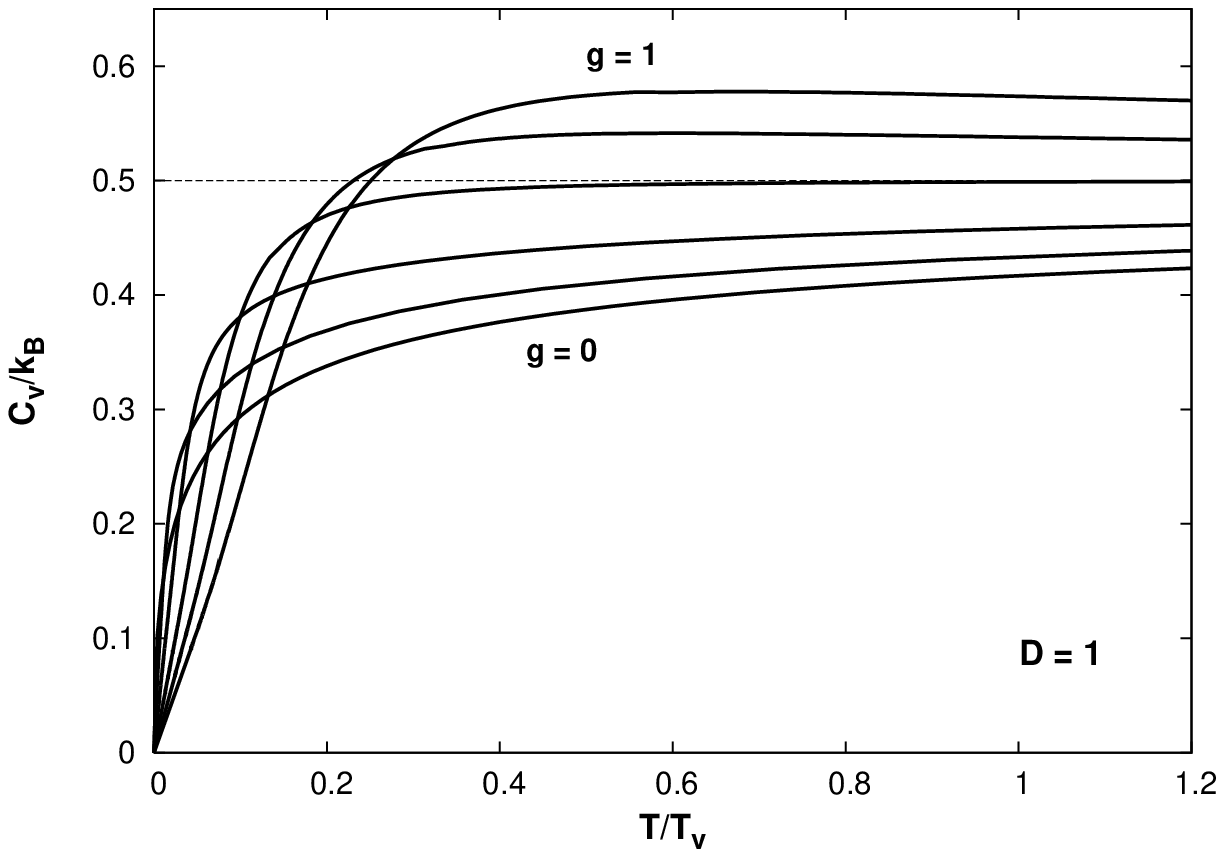}
  \end{minipage}%
  \begin{minipage}[c]{0.5\textwidth}
     \centering \includegraphics[width=85mm]{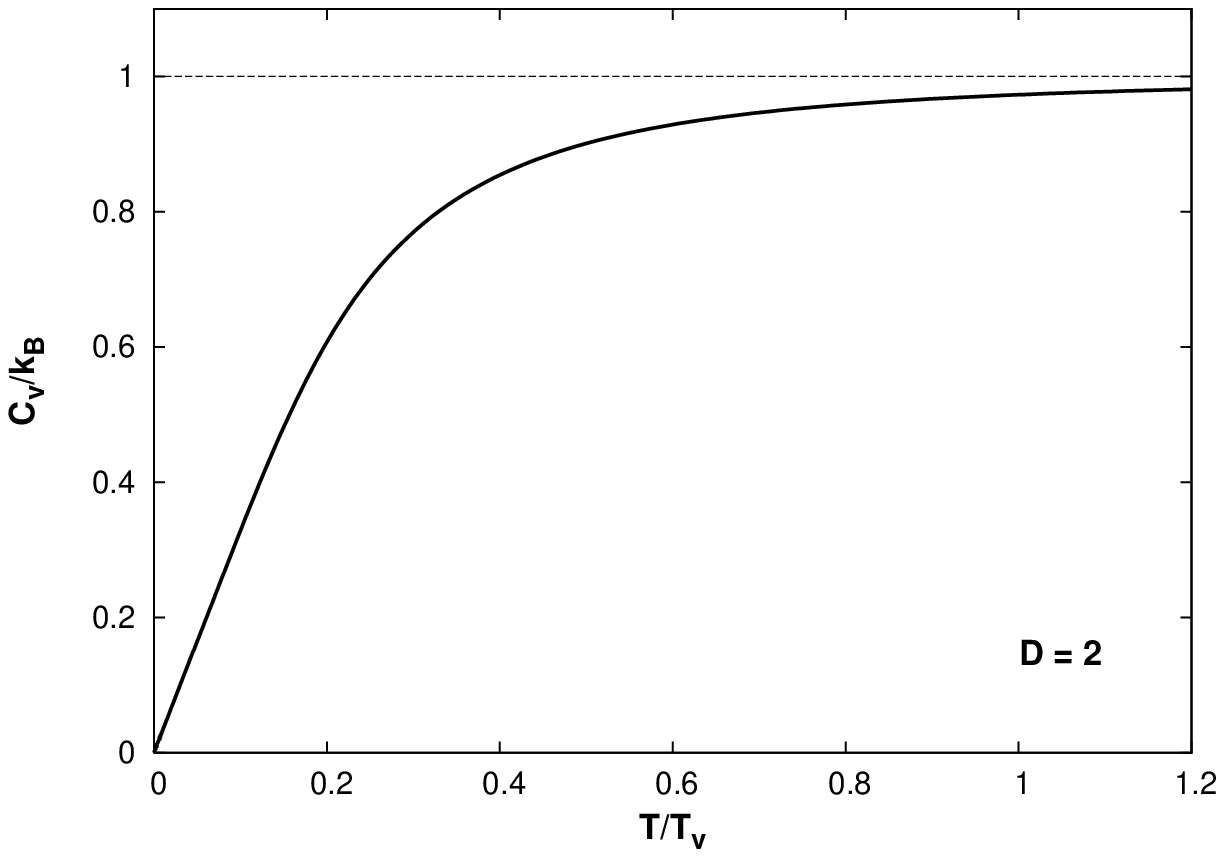}
  \end{minipage}\\
   \begin{minipage}[c]{.5\textwidth}
     \centering \includegraphics[width=85mm]{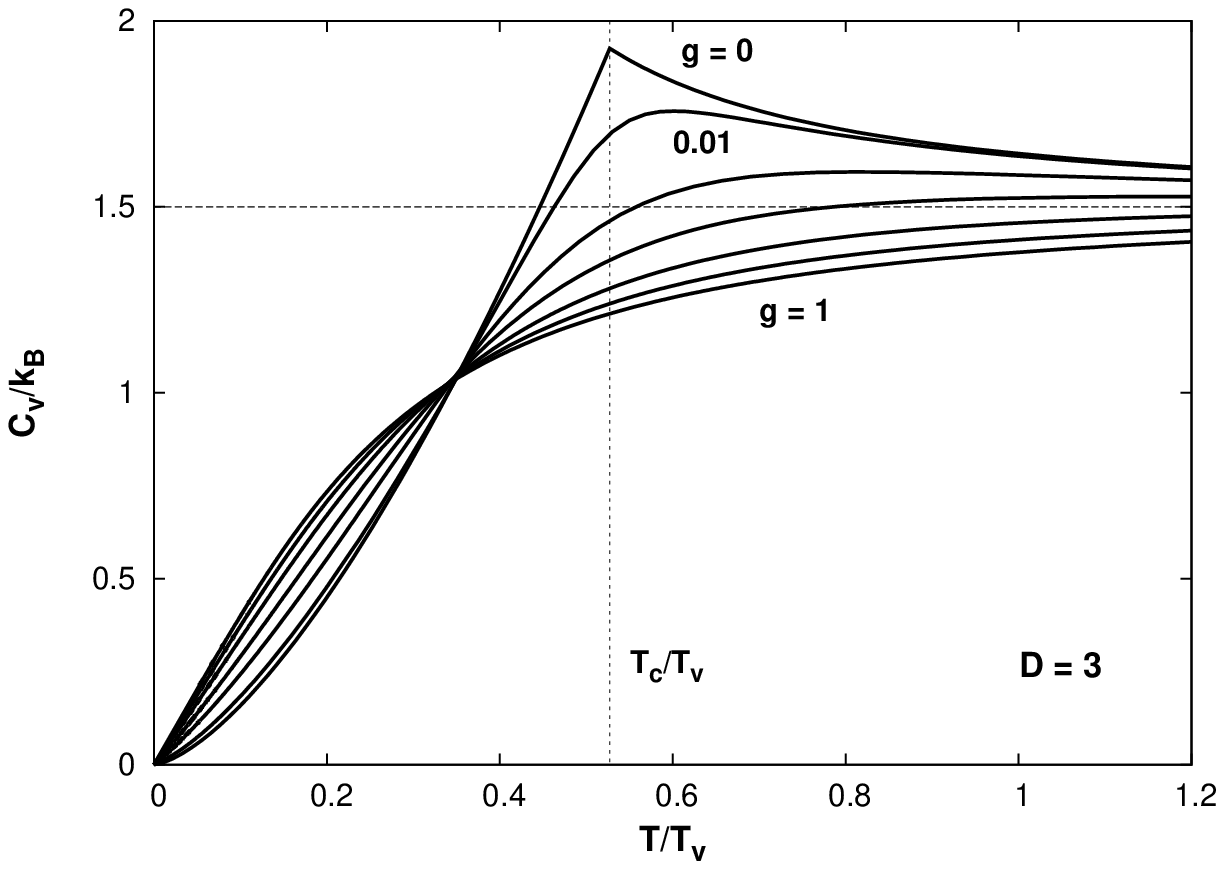}
  \end{minipage}%
  \begin{minipage}[c]{.5\textwidth}
     \centering \includegraphics[width=85mm]{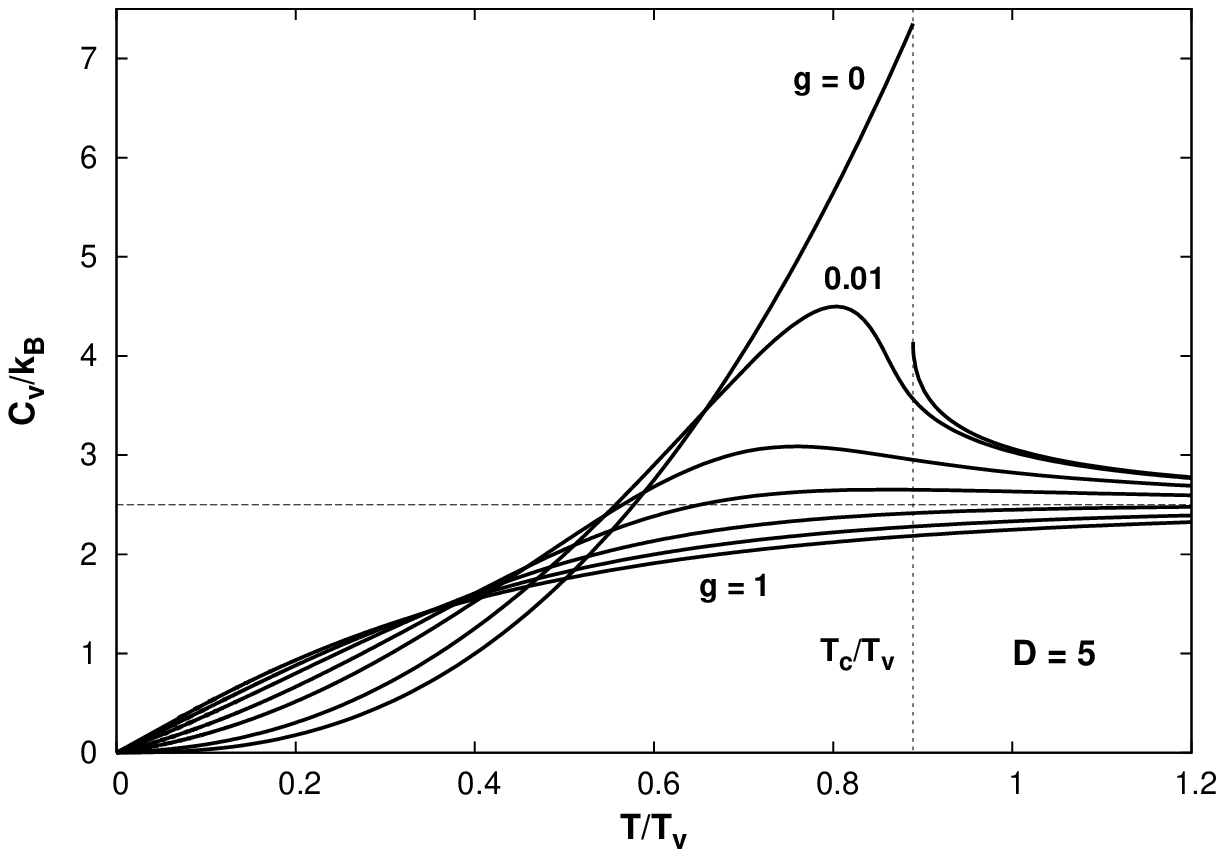}
  \end{minipage}
  \caption{Heat capacity in $\mathcal{D} = 1,2,\ldots,5$ for $g=0$, $0.01$, $0.1$,
    $0.25$, $0.5$, $0.75$, $1$ (from top down or bottom up). In $\mathcal{D}=1$
    the curves for $g=0$, $0.01$ are unresolved. In $\mathcal{D}=2$ the heat
    capacity is independent of $g$. In $\mathcal{D}=5$ the heat capacity has a
    discontinuity $\Delta C_v/k_B\simeq 3.21$ at $T_c$. Note the different vertical scales.}
  \label{fig:gencsCv}
\end{figure}
%%%%%%%%%%%%%%%%%%%%%%%%%%%%%%%%%%%%%%%%%%%%%
\end{widetext}

%
%%%%%%%%%%%%%%%%%%%%%%%%%%%%%%%%%%%%%%%%%%%%%%%
From (\ref{eq:52}) and (\ref{eq:51}) we know that the scaled internal energy 
$U/ \mathcal{N}k_BT_v$ differs from the isochore (\ref{eq:111}) by a mere overall factor
$\mathcal{D}/2$. The isochoric heat capacity, $C_v\doteq\mathcal{N}^{-1}(\partial U/ \partial
T)_v$, derived from that expression reads
\begin{equation}
  \label{eq:94lj1a}
   \frac{C_v}{k_B} = \frac{\mathcal{D}}{2}\left[\left(\frac{\mathcal{D}}{2} + 1\right)
  \frac{G_{\mathcal{D}/2+1}(z,g)}{G_{\mathcal{D}/2}(z,g)} - 
  \frac{\mathcal{D}}{2}\frac{G_{\mathcal{D}/2}(z,g)} 
{G_{\mathcal{D}/2-1}(z,g)}\right]. 
\end{equation}
In Fig. \ref{fig:gencsCv} we show plots of
$C_v/k_B$ versus $T/T_v$ from (\ref{eq:111}) for various $g$ and
$\mathcal{D}$.
In $\mathcal{D}=2$ the heat capacity is
independent of $g$, a result known since 1964 for the FD and BE
gases and recently extended to $0<g<1$
\cite{May64,VRH95,Lee97,SC04,Angh02}:

\begin{align}
     \frac{C_v}{k_B} = 
     \frac{\pi^{2}}{3} \frac{T}{T_{v}} - \frac{T_{v}}{T}
     \sum_{n=1}^{\infty}
     \left(1+\frac{2}{n}\frac{T}{T_{v}} +\frac{2}{n^{2}}\frac{T^{2}}{T_{v}^{2}}\right)e^{-nT_{v}/T}
\end{align}

In $\mathcal{D}<2$ only the curves for $g>\frac{1}{2}$ have a local maximum and
in $\mathcal{D}>2$ only the curves for $g<\frac{1}{2}$. This double switch is
reflected in the leading correction to the MB behavior at high $T$:
  \begin{equation}
    \label{eq:132}
    \frac{C_v}{k_B} \stackrel{T\to\infty}{\leadsto} 
    \frac{\mathcal{D}}{2}\left[
      1+\frac{(1/2-g)(\mathcal{D}/2-1)}{2^{\mathcal{D}/2}}
      \left(\frac{T_v}{T}\right)^{\mathcal{D}/2}
    \right].
  \end{equation}
The leading low-$T$ behavior of $C_v$ is linear for all $g>0$:
\begin{equation}
  \label{eq:133}
  \frac{C_v}{k_B} \stackrel{T\to0}{\leadsto} 
  \frac{\pi^2}{6}
  \frac{\mathcal{D}g^{(\mathcal{D}-2)/ \mathcal{D}}}{[\Gamma(\mathcal{D}/2+1)]^{2/ \mathcal{D}}}
    \frac{T}{T_v}.
\end{equation}
In $\mathcal{D}<2~(\mathcal{D}>2)$, the initial slope increases (decreases) with
decreasing $g$, in agreement with the numerical results of Fig. \ref{fig:gencsCv}.

The well-known power-law behavior of the bosonic heat capacity at low $T$ reads
\begin{equation}
  \label{eq:134}
  \frac{C_v}{k_B} \stackrel{T\to0}{\leadsto} 
  \frac{\mathcal{D}}{2}\left(\frac{\mathcal{D}}{2}+ 1\right)
  \zeta\left(\frac{\mathcal{D}}{2}+1\right)
  \left(\frac{T}{T_v}\right)^{\mathcal{D}/2}.
\end{equation}
It represents the leading singularity of the exact result in $\mathcal{D}\leq2$
and, at the same time, the exact result itself for $0\leq T\leq T_c$ in
$\mathcal{D}>2$. The latter result can be rewritten in the form
\begin{equation}
  \label{eq:135}
 \frac{C_v}{k_B}= \frac{\mathcal{D}}{2}\left(\frac{\mathcal{D}}{2}+1\right)
  \frac{\zeta\left(\mathcal{D}/2+1\right)}{\zeta\left(\mathcal{D}/2\right)} 
\left(\frac{T}{T_c}\right)^{\mathcal{D}/2}
\end{equation}
for $0\leq T\leq T_c$ with $T_c$ from (\ref{eq:112}).

Inspection of the bosonic heat capacity results (\ref{eq:94lj1a}) and
(\ref{eq:135}) for $T\to T_c$ from above and below, respectively, in
$\mathcal{D}>2$ shows a qualitative change of behavior in $\mathcal{D}=4$. In
$2<\mathcal{D}\leq4$ the heat capacity $C_v$ has a local maximum at $T_c$
with a discontinuous slope. In $\mathcal{D}>4$ the function
itself becomes discontinuous at $T_c$ as is evident in Fig.~\ref{fig:gencsCv}.
The size of the discontinuity is
  \begin{equation}
    \label{eq:136}
    \frac{\Delta C_v}{k_B}= \frac{\mathcal{D}^2}{4}
\frac{\zeta\left(\mathcal{D}/2\right)}{\zeta\left(\mathcal{D}/2-1\right)}
\qquad (\mathcal{D}>4).
\end{equation}
There is no latent heat. Only in the limit $\mathcal{D}\to\infty$ a
first-order transition emerges. In large $\mathcal{D}$ the heat capacity
stays very close to zero from $T=0$ up to the near vicinity of $T_c$, where
it shoots up to a high value of O$(\mathcal{D}^2)$. At $T_c$ it drops back to
a value of O$(\mathcal{D})$ and remains nearly constant. In the limit
$\mathcal{D}\to\infty$ this spike turns into a delta function and its weight
determines the latent heat.

\begin{widetext}
%%%%%%%%%%%%%%%%%%%%%%%%%%%%%%%%%%%%%%%%%%%%
\begin{figure}[h!]
  \centering
  \begin{minipage}[c]{0.5\textwidth}
     \centering \includegraphics[width=85mm]{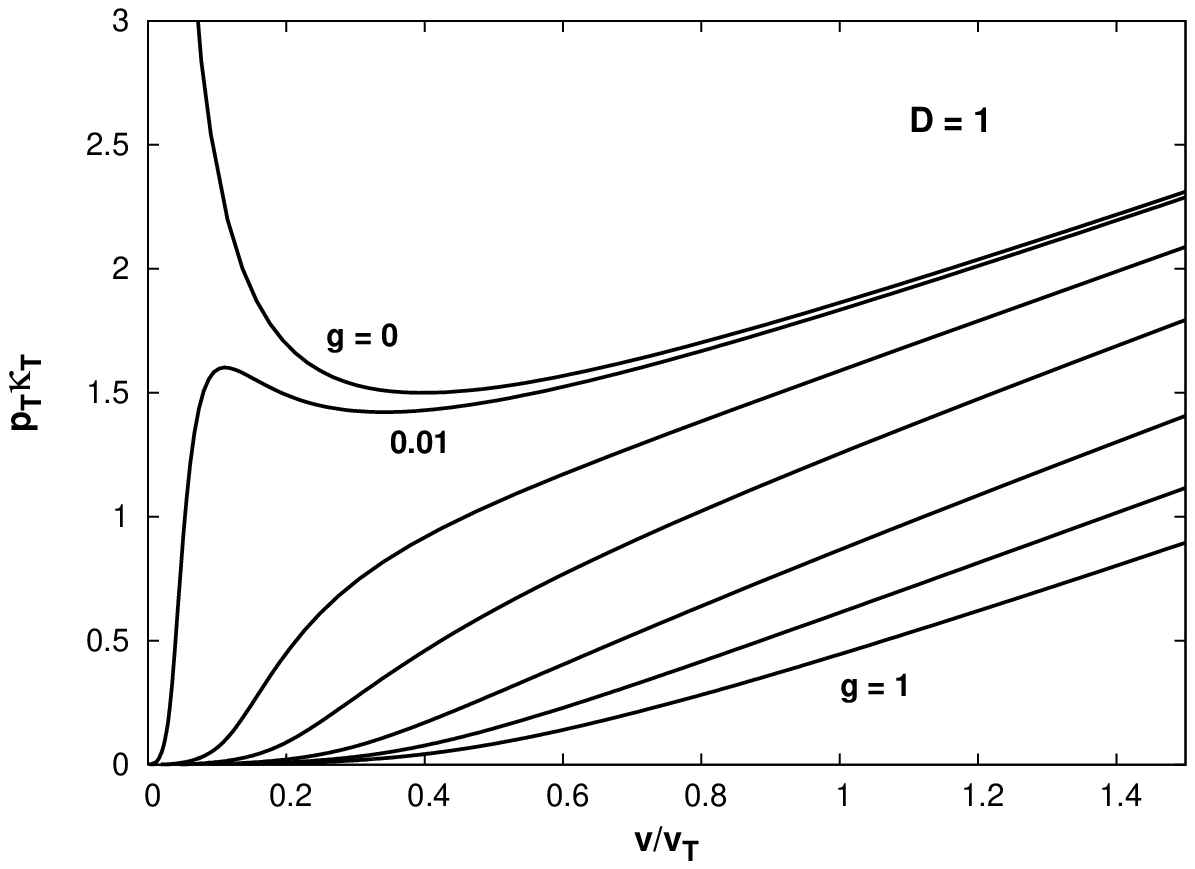}
  \end{minipage}%
  \begin{minipage}[c]{0.5\textwidth}
     \centering \includegraphics[width=85mm]{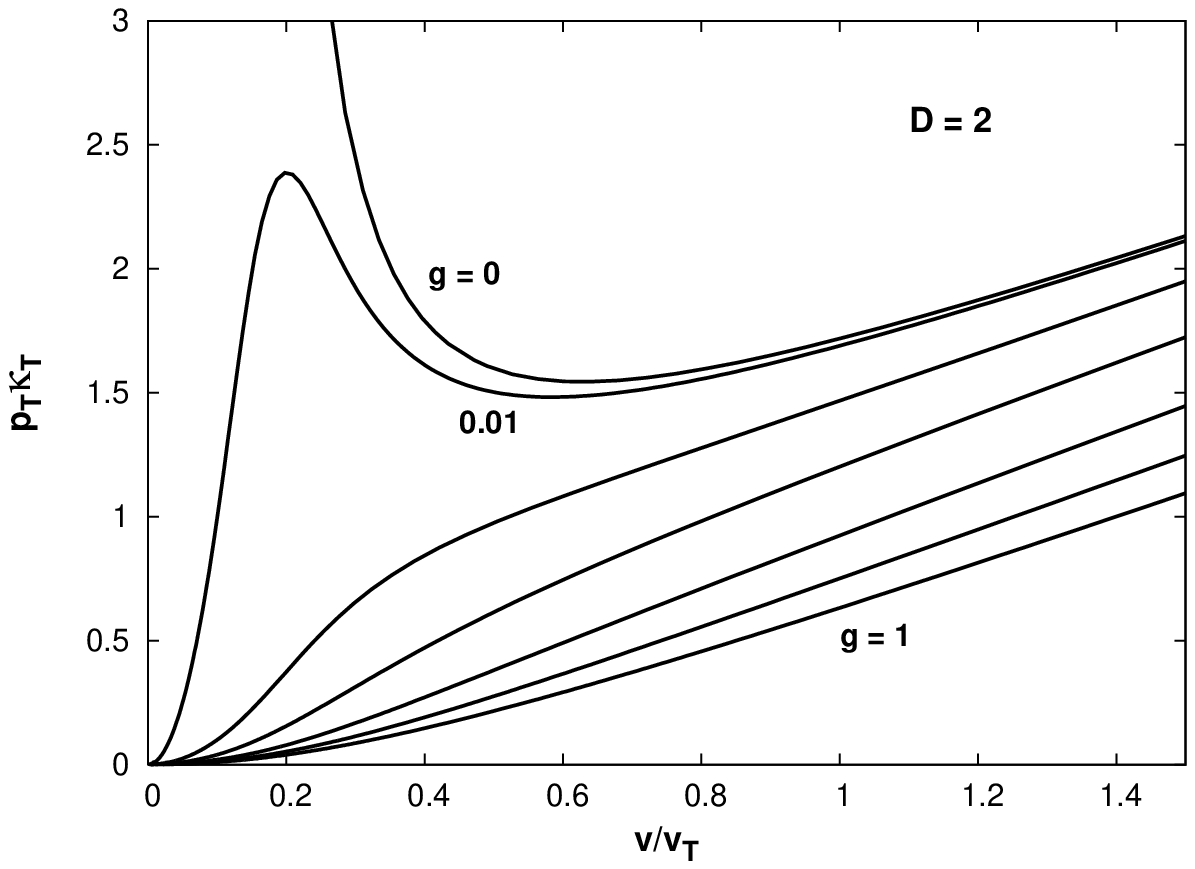}
  \end{minipage}\\
   \begin{minipage}[c]{.5\textwidth}
     \centering \includegraphics[width=85mm]{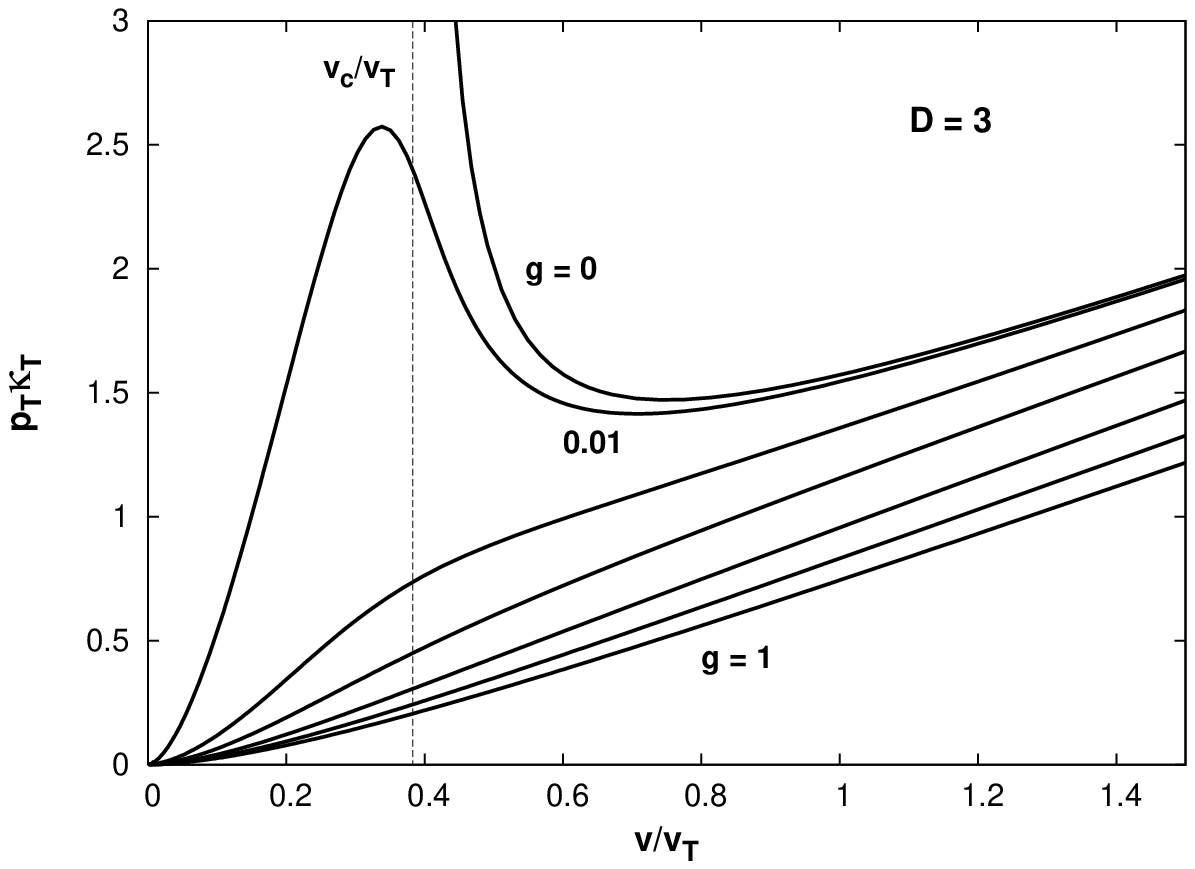}
  \end{minipage}%
  \begin{minipage}[c]{.5\textwidth}
     \centering \includegraphics[width=85mm]{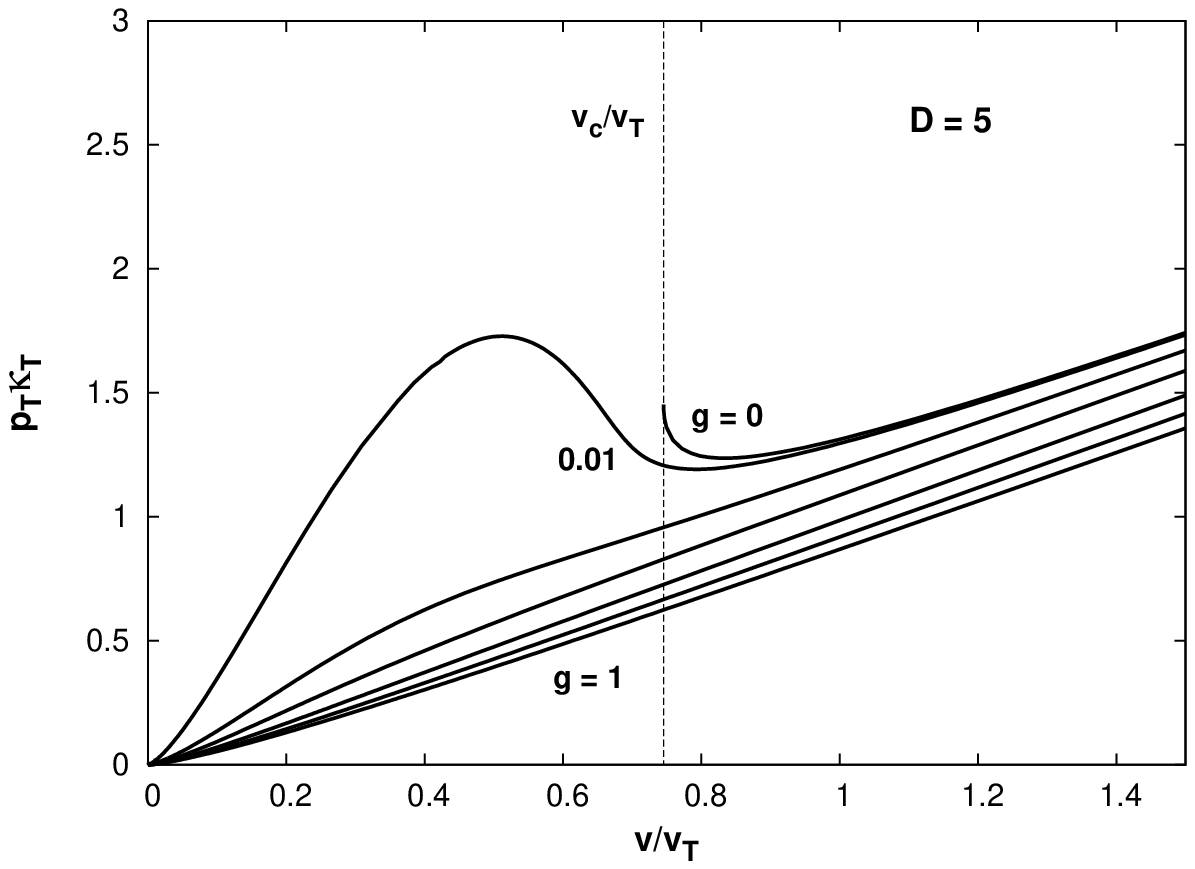}
  \end{minipage}
  \caption{Isothermal compressibility in $\mathcal{D} = 1,2,3,5$ for $g=0$,
    $0.01$, $0.1$, $0.25$, $0.5$, $0.75$, $1$ (from top down). The endpoint
    value of the bosonic curve in $\mathcal{D}=5$ is $p_T\kappa_T\simeq1.45$.}
  \label{fig:gencsistcmp}
\end{figure}
%%%%%%%%%%%%%%%%%%%%%%%%%%%%%%%%%%%%%%%%%%%%%
\end{widetext}

%%%%%%%%%%%%%%%%%%%%%%%%%%%%%%%%%%%%%%%%%%%%%%%
%
\subsection{Isothermal compressibility}\label{sec:isothcomp}  
%
%%%%%%%%%%%%%%%%%%%%%%%%%%%%%%%%%%%%%%%%%%%%%%%
We obtain the following parametric representation of $\kappa_T\doteq -v^{-1}(\partial v/ \partial
p)_T$ from (\ref{eq:114a}):
\begin{equation}
  \label{eq:137}
  p_T\kappa_T= \frac{v}{v_T}\,\frac{G_{\mathcal{D}/2-1}(z,g)}%
{G_{\mathcal{D}/2}(z,g)},
\end{equation}
with $v/v_T$ from (\ref{eq:114a}). In Fig.~\ref{fig:gencsistcmp} we plot
$p_T\kappa_T$ versus $v/v_T$ for various $g$ and
$\mathcal{D}$.

The leading correction to MB behavior at low density, $v/v_T\gg1$,
\begin{equation}
    \label{eq:139}
    p_T\kappa_T \stackrel{v\to\infty}{\leadsto} \frac{v}{v_T}
    \left[1+\frac{1/2-g}{2^{\mathcal{D}/2-1}}\frac{v_T}{v}\right],
\end{equation}
is strongest in low $\mathcal{D}$.  The compressibility curves for
$g>\frac{1}{2}$ display fermion-like behavior across the entire range of
$v/v_T$. The value of $\kappa_T$ stays below the MB value and decreases monotonically
with decreasing $v/v_T$, approaching zero in the limit $v\to0$, reflecting the
repulsive core of the statistical interaction. For $g<\frac{1}{2}$, by contrast,
the curves start out above the MB line. Here the long-range attractive part of
the statistical interaction is dominant, producing boson-like behavior. As the
particle density increases a crossover from boson-like behavior at low density
to fermion-like behavior at high density manifests itself as a shoulder or as a
precipitous drop after a smooth local maximum.  The leading high-density term of all
compressibility curves for $g>0$ is a power-law with $\mathcal{D}$-dependent
exponent:
\begin{equation}
  \label{eq:5}
  p_T\kappa_T \stackrel{v \to 0}{\leadsto} 
  \frac{\mathcal{D}}{2g^{2/ \mathcal{D}}[\Gamma(\mathcal{D}/2+1)]^{2/ \mathcal{D}}}\,
  \left(\frac{v}{v_T}\right)^{\frac{\mathcal{D}+2}{\mathcal{D}}}. 
\end{equation}

The uppermost curve in each panel of Fig.~\ref{fig:gencsistcmp} is for bosons.
It shares with all the other curves a decreasing initial trend as $v/v_T$ is
lowered in the low-density regime. Then it goes through a smooth minimum (a
property shared with curves for $g\ll1$) and rises to either a divergence or a
cusp at $v_c$.  In $\mathcal{D}\leq2$ we have $v_c=0$. Here the isothermal
compressibility exhibits a power-law divergence, $\kappa_T\sim
(v/v_T)^{-\mathcal{D}/(2-\mathcal{D})}$, in $1\leq\mathcal{D}<2$, and an
exponential divergence, $\kappa_T\sim (v/v_T)^2e^{v/v_T}$, in $\mathcal{D}=2$. In
$2<\mathcal{D}\leq4$ we have $v_c>0$ as given in (\ref{eq:115a}) and $\kappa_T$ still
diverges. However, in $\mathcal{D}>4$ the divergence turns into a cusp at
$p_{Tc}\kappa_{Tc} = \zeta(\mathcal{D}/2-1)/ \left[\zeta(\mathcal{D}/2)\right]^2$.  At
$v<v_c$ we have $\kappa_T=\infty$.

\begin{widetext}
%%%%%%%%%%%%%%%%%%%%%%%%%%%%%%%%%%%%%%%%%%%%
\begin{figure*}[t!]
  \centering
  \begin{minipage}[c]{0.5\textwidth}
     \centering \includegraphics[width=85mm]{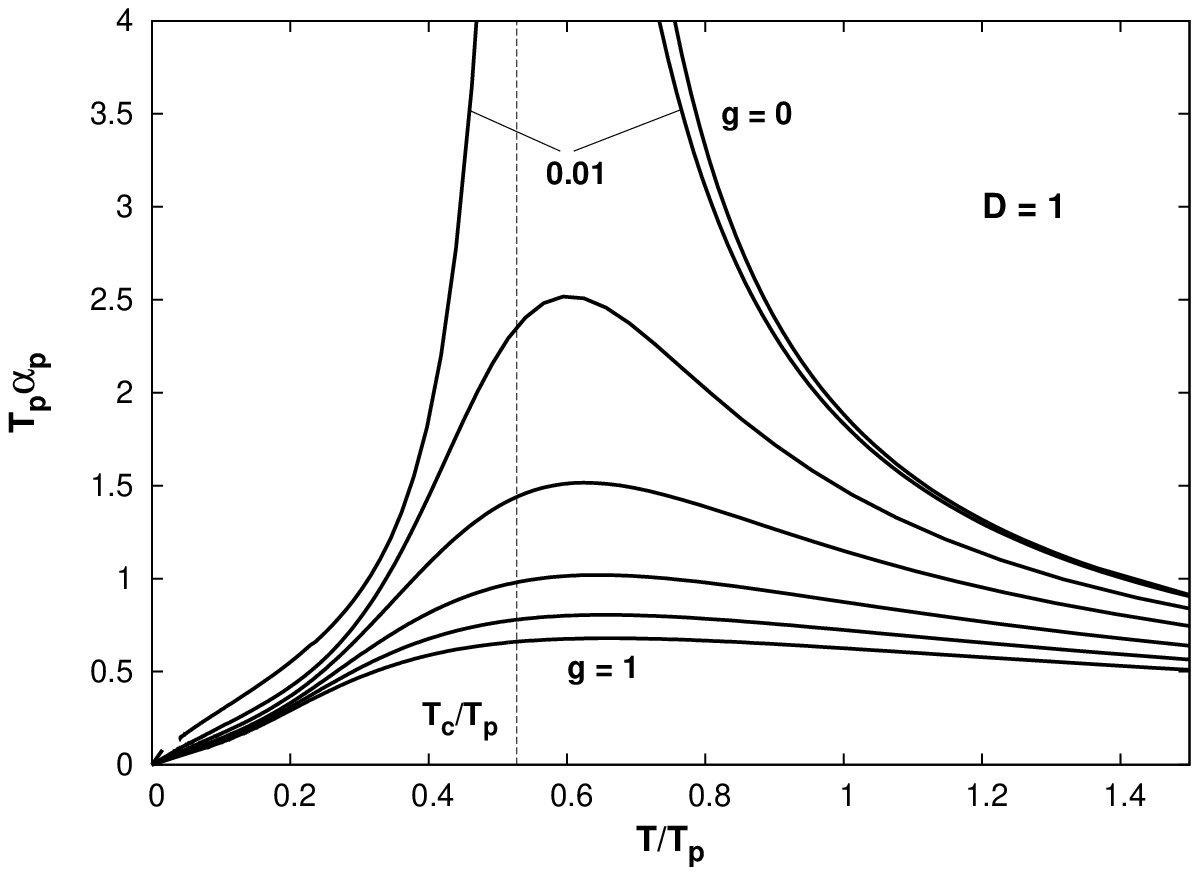}
  \end{minipage}%
  \begin{minipage}[c]{0.5\textwidth}
     \centering \includegraphics[width=85mm]{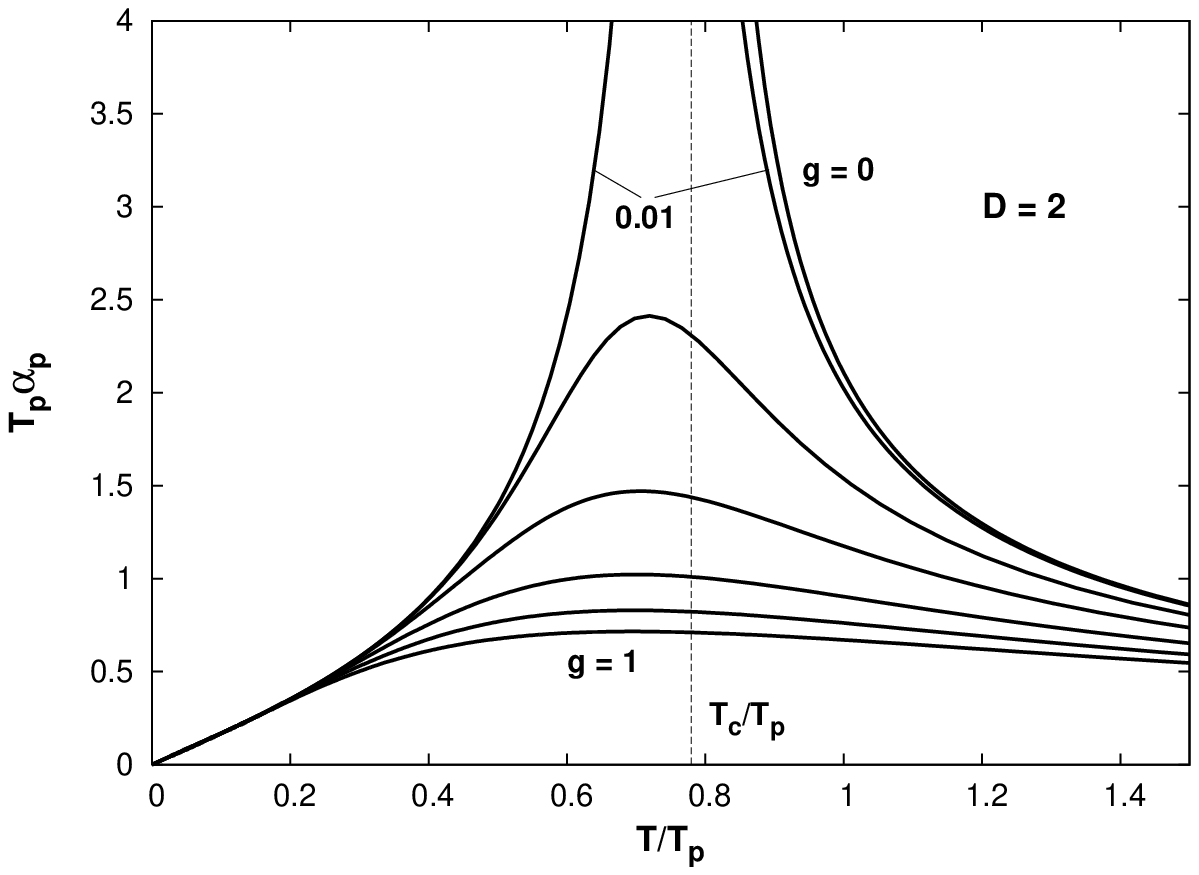}
  \end{minipage}\\
   \begin{minipage}[c]{.5\textwidth}
     \centering \includegraphics[width=85mm]{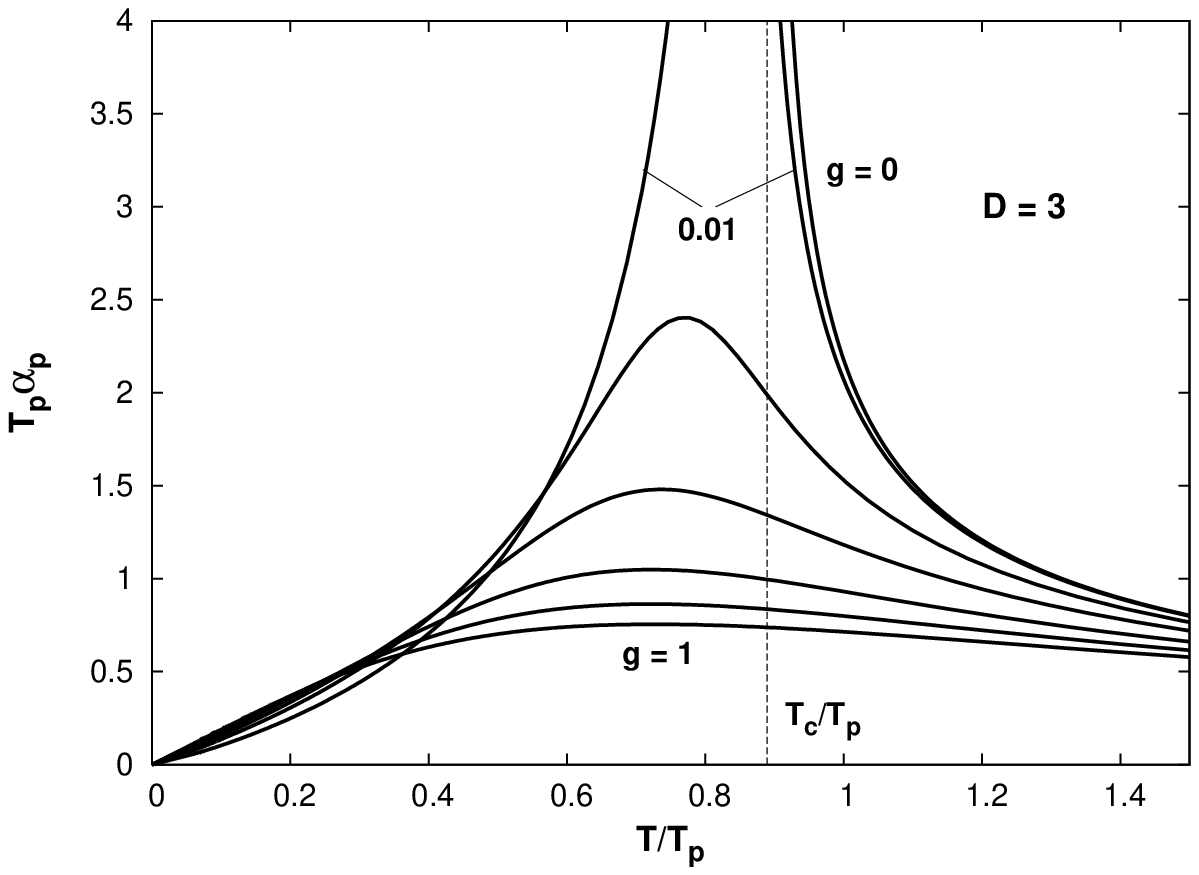}
  \end{minipage}%
  \begin{minipage}[c]{.5\textwidth}
     \centering \includegraphics[width=85mm]{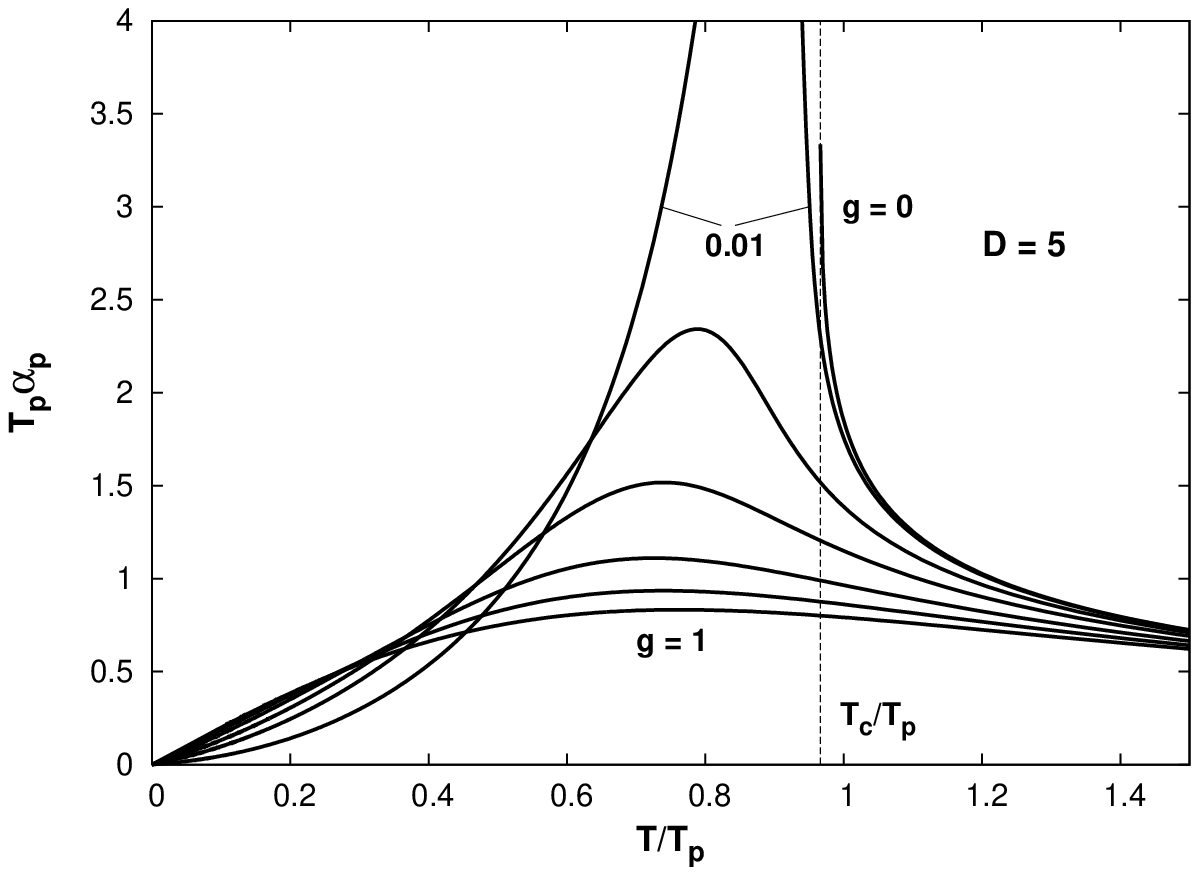}
  \end{minipage}
   \caption{Isobaric expansivity in $\mathcal{D} = 1,2,3,5$ for $g=0$,
     $0.01$, $0.1$, $0.25$, $0.5$, $0.75$, $1$ (from top down almost
     everywhere). The smooth maximum for $g=0.01$ has the value $T_p\alpha_p\simeq8.77$,
     $9.69$, $9.49$, $6.84$ in $\mathcal{D} = 1,2,3,5$, respectively. The endpoint
     value of the bosonic curve in $\mathcal{D}=5$ is $T_p\alpha_p\simeq3.34$.}
   \label{fig:gencsnewisbexp}
\end{figure*}
%%%%%%%%%%%%%%%%%%%%%%%%%%%%%%%%%%%%%%%%%%%%%
\end{widetext}

%%%%%%%%%%%%%%%%%%%%%%%%%%%%%%%%%%%%%%%%%%%%%%%
%
\subsection{Isobaric expansivity}\label{sec:isobexp}  
%
%%%%%%%%%%%%%%%%%%%%%%%%%%%%%%%%%%%%%%%%%%%%%%%
We calculate $\alpha_p\doteq v^{-1}(\partial v/ \partial T)_p$ from the general relation
$\alpha_p=\kappa_T(\partial p/ \partial T)_v$ and the CS-specific relation $C_v=(\mathcal{D}/2)v(\partial
p/ \partial T)_v$ to arrive at
\begin{align}
  \label{eq:138}
  T_p\alpha_p = \frac{T_p}{T}\left[\left(\frac{\mathcal{D}}{2}+1\right)
    \frac{G_{\mathcal{D}/2+1}(z,g)G_{\mathcal{D}/2-1}(z,g)}{G_{\mathcal{D}/2}(z,g)
G_{\mathcal{D}/2}(z,g)} -\frac{\mathcal{D}}{2}\right]
\end{align}
with $T/T_p$ from (\ref{eq:27}). In Fig.~\ref{fig:gencsnewisbexp} we show
expansivity curves $T_p\alpha_p$ versus $T/T_p$ for various $g$ and $\mathcal{D}$.
We observe that the correction to MB behavior at high $T$ is such that the expansivity is
suppressed for $g>\frac{1}{2}$ and enhanced for $g<\frac{1}{2}$:
\begin{equation}
    \label{eq:138highT}
    T_p\alpha_p \stackrel{T\to\infty}{\leadsto} \frac{T_p}{T}
    \left[1+\frac{(1/2-g)(\mathcal{D}/2+1)}{2^{\mathcal{D}/2}}
    \left(\frac{T_p}{T}\right)^{\mathcal{D}/2+1}\right].
\end{equation} 

The characteristic features of all compressibility curves for $0<g<1$ are a
smooth maximum and a linear approach to zero in the low-$T$ limit. The maximum
is flat near the fermion limit and becomes increasingly high and narrow as the
boson limit is approached. For $g\ll1$ the position of the maximum is close to
the bosonic $T_c$ as given in (\ref{eq:117}).  The slope of the linear low-$T$
behavior of $\alpha_p$ depends on $g$ and $\mathcal{D}$:
\begin{equation}
  \label{eq:8}
%  T_p\alpha_p \stackrel{T\to0}{\leadsto}
%  \frac{T}{T_p}\frac{\mathcal{D}}{2}\,g^{(\mathcal{D}-2)/(\mathcal{D}+2)}\,
%  \frac{\pi^2}{3}[\Gamma(\mathcal{D}/2+1)]^{-4/(\mathcal{D}+2)}.
%  \\ 
  T_p\alpha_p \stackrel{T\to0}{\leadsto}
  \frac{\pi^2}{6} 
  \frac{\mathcal{D}g^{(\mathcal{D}-2)/(\mathcal{D}+2)}}{[\Gamma(\mathcal{D}/2+1)]^{4/(\mathcal{D}+2)}}
  \frac{T}{T_p} .
\end{equation}
Note the absence of any $g$-dependence in $\mathcal{D}=2$ and the opposite
trends regarding $g$-dependence in $\mathcal{D}<2$ and $\mathcal{D}>2$
reminiscent of trends seen in the heat capacity.

The bosonic expansivity increases monotonically with decreasing $T$ and ends in
a singularity as $T_c$ is approached from above. In $\mathcal{D}\leq4$ the
singularity is a divergence and in $\mathcal{D}>4$ it is a cusp.
%\newpage

%%%%%%%%%%%%%%%%%%%%%%%%%%%%%%%%%%%%%%%%%%%%%%%
%
\subsection{Velocity of sound}\label{sec:velsou}  
%
%%%%%%%%%%%%%%%%%%%%%%%%%%%%%%%%%%%%%%%%%%%%%%%
We start from the relation $c=(\rho\kappa_S)^{-1/2}$ for the velocity of sound, where
$\rho=m/v$ is the mass density and $\kappa_S$ the adiabatic compressibility. We use
standard relations between response functions to arrive at the following
expression, all in terms of dimensionless entities previously
determined \footnote{For consistency, we must use reference values $T_p$,
  $T_v$, $v_T$ derived from the first expression (\ref{eq:10}) of $\lambda_T$ in
  this context.}:
\begin{equation}
  \label{eq:11}
   \frac{mc^2}{k_BT}= \frac{(v/v_T)}{(p_T\kappa_T)}\left[1+ 
    \frac{(T/T_p)^2(v/v_T)(T_p\alpha_p)^2}{(p_T\kappa_T)(C_v/k_B)}\right]. 
\end{equation}
For the CS model the right-hand side of (\ref{eq:11}) can be greatly simplified:
\begin{equation}
  \label{eq:18}
  \frac{mc^2}{k_BT} =
  \gamma\,\frac{G_{\mathcal{D}/2+1}(z,g)}{G_{\mathcal{D}/2}(z,g)},\qquad \gamma\doteq1+\frac{2}{\mathcal{D}}. 
\end{equation}
How does the velocity of sound $c$ vary with temperature $T$ if we keep the
(average) particle density $1/v$ or the (average) pressure $p$ fixed?
Remarkably, we find very simple universal relations between $c$ and the
isochore or isobar itself for the two situations, respectively:
\begin{equation}
  \label{eq:20}
  \frac{mc^2}{k_BT_v} =\gamma\,\frac{p}{p_v}\qquad (v=\mathrm{const}.),
\end{equation}
\begin{equation}
  \label{eq:21}
   \frac{mc^2}{k_BT_p} =\gamma\,\frac{v}{v_p}\qquad (p=\mathrm{const}.),
\end{equation}
with the dependence of $p/p_v$ on $T/T_v$ as discussed in Sec.~\ref{sec:isoc}
and the dependence of $v/v_p$ on $T/T_p$ as discussed in Sec.~\ref{sec:isob}.

We conclude that $c$ has a monotonic dependence on $T$ no matter whether
we keep $v$ or $p$ constant. for $g>0$ the velocity of sound stays nonzero in
the limit $T\to0$. In the boson gas $c$ is affected by the onset of BEC. There
are no sound waves in the condensate. The isochores at $T<T_c$ in
Fig.~\ref{fig:gencsppv} or the vertical portions of the isobars at $T=T_c$ in
Fig.~\ref{fig:gencsisob} give us information about $c$ in the gas coexisting
with the condensate.

%\pagebreak
%%%%%%%%%%%%%%%%%%%%%%%%%%%%%%%%%%%%%%%%%%%%%%%
%
\section{Effects of soft trap walls}\label{sec:eoscw}  
%
%%%%%%%%%%%%%%%%%%%%%%%%%%%%%%%%%%%%%%%%%%%%%%%
Consider an ideal quantum gas in $\mathcal{D}\geq1$, trapped by a
spatially isotropic power-law potential \cite{BPK87,BK91}
\begin{equation} 
  \label{eq:1lj2a} 
  \mathcal{U}(r)\doteq \mathcal{U}_0\left(\frac{r}{R}\right)^\eta, 
\end{equation} 
where $r$ is the radial coordinate in $\mathcal{D}$-dimensional space. The size
of the trap is determined by the ``width'' $R$ and the ``depth''
$\mathcal{U}_0$ of the potential well. The softness of the confinement is
controlled by the exponent $\eta$. Harmonic traps have $\eta=2$. Lowering the value
of $\eta$ makes the surrounding wall softer, increasing it makes the wall harder.
For $\eta\to\infty$, the wall becomes rigid and the bottom of the potential becomes
flat, which corresponds to the situation considered in Sec.~\ref{sec:gcs}.

Softening the trap walls affects the density
of energy levels. For the power-law potential (\ref{eq:1lj2a}) expression
(\ref{eq:22}) must be replaced by
\begin{equation} 
  \label{eq:7lj2a} 
  D(\epsilon_0) = 
  \frac{\mathcal{V}}{2^\mathcal{D}\pi^{\mathcal{D}/2}\Gamma(\mathcal{D}/2)}  
\,\epsilon_0^{\mathcal{D}/2-1}\left(\frac{\epsilon_0}{\mathcal{U}_0}\right)^{\mathcal{D}/ 
  \eta}Q_\mathcal{D}(\eta)
\end{equation} 
with 
\begin{equation} 
  \label{eq:2blj2a} 
  \mathcal{V} =
  \frac{\pi^{\mathcal{D}/2}}{\Gamma(\mathcal{D}/2+1)}R^{\mathcal{D}},\quad 
  Q_\mathcal{D}(\eta)\doteq \frac{\Gamma(\mathcal{D}/ \eta+1)\Gamma(\mathcal{D}/2)}  
{\Gamma(\mathcal{D}/ \eta+\mathcal{D}/2)}. 
\end{equation} 
In soft-wall traps it makes sense to consider processes at
$\mathcal{V}=\mathrm{const}$ instead of isochoric processes. Under these
circumstances the gas expands when heated up.
 In generalization of relations (\ref{eq:52}) and
(\ref{eq:51}) we now have
\begin{equation} 
  \label{eq:9blj2a} 
  \hspace*{-5mm}\mathcal{N}=\frac{\mathcal{V}}{\lambda_T^{\mathcal{D}}}\Gamma(\mathcal{D}/\eta+1) 
  \left(\frac{k_B}{\mathcal{U}_0}\right)^{\mathcal{D}/\eta} 
  G_{\mathcal{D}/2+\mathcal{D}/\eta}(z,g),
\end{equation} 
\begin{align} 
  \nonumber
  U = &  
  \frac{\mathcal{V}k_BT}{\lambda_T^{\mathcal{D}}}\left(\frac{k_BT}{\mathcal{U}_0}\right)^{\mathcal{D}/\eta}  
   \\   \label{eq:ulj2a}  & \times 
  \Gamma\left(\frac{\mathcal{D}}{\eta}+1\right)\left(\frac{\mathcal{D}}{\eta} + \frac{\mathcal{D}}{2}\right) 
  G_{\mathcal{D}/2+\mathcal{D}/\eta+1}(z,g). 
\end{align} 
As before, an additive term $z/(1-z)$ has to be considered in (\ref{eq:9blj2a}) if $g=0$. We
now use the reference temperature 
\begin{align} 
  \nonumber
  T_v = & \left(\frac{\mathcal{N}}{\mathcal{V}}\right)^{\frac{1}{\mathcal{D}/\eta + \mathcal{D}/2}} 
  \left(\frac{4\pi}{k_B}\right)^{\frac{\mathcal{D}/2}{\mathcal{D}/\eta +
      \mathcal{D}/2}} 
  \left(\frac{\mathcal{U}_0}{k_B}\right)^{\frac{\mathcal{D}/\eta}{\mathcal{D}/\eta + \mathcal{D}/2}} 
  \\   \label{eq:ulj2b} & \times 
  \left[\Gamma(\mathcal{D}/\eta + 1)\right]^{\frac{-1}{\mathcal{D}/\eta + \mathcal{D}/2}}. 
\end{align} 
For bosons  the onset of BEC occurs at $T_c>0$ in
$\mathcal{D}>\mathcal{D}_\mathrm{m}=2\eta/(2+\eta)$. The marginal
dimensionality decreases from $\mathcal{D}_\mathrm{m}=2$ to
$\mathcal{D}_\mathrm{m}=1$ as the rigid-wall container softens into a
harmonic trap. The transition
temperature in units of $T_v$ is
\begin{equation} 
  \label{tct0lj2a} 
  \frac{T_c}{T_v} = \left[\zeta(\mathcal{D}/\eta + 
    \mathcal{D}/2)\right]^{\frac{-1}{\mathcal{D}/\eta + \mathcal{D}/2}}.  
\end{equation} 

%%%%%%%%%%%%%%%%%%%%%%%%%%%%%%%%%%%%%%%%%%%%%%%
%
\section{Conclusion and outlook}\label{sec:concl}  
%
%%%%%%%%%%%%%%%%%%%%%%%%%%%%%%%%%%%%%%%%%%%%%%%
We have explored the thermodynamics of the generalized CS model in
$\mathcal{D}\geq1$ dimensions. In this model the statistical interaction is limited
to pairs of particles with identical momenta. This reduces the coupling,
effectively, to a statistical exclusion condition. Several observed phenomena
suggest the presence of a long-range attractive part and a short-range repulsive
part in this particular statistical interaction.

The generalized CS model is found to preserve several attributes that are
characteristic of the familiar ideal BE and FD gases -- attributes which may
very well count as hallmarks of ideal quantum gases in general: (i) The average
level occupancy $\langle n(\epsilon_0)\rangle$ is a unique function of $(\epsilon_0-\mu)/k_BT$ for
given $g$, independent of $\mathcal{D}$. (ii) The fundamental thermodynamic
relations (\ref{eq:50})--(\ref{eq:51}) for given $g$ and with the prescribed
structure on the left-hand side are unique functions of the fugacity $z$.
(iii) The right-hand sides of Eqs. (\ref{eq:50}) and (\ref{eq:51}) are
identical.

The consequences include that the isochore and the (properly scaled) internal
energy have the same dependence on $T$, that the quantity $pV/ \mathcal{N}k_BT$
has a unique dependence on $z$ for given $\mathcal{D}$ and that the square of
the (properly scaled) velocity of sound at constant $v$ (constant $p$) has the
same $T$-dependence as the isochore (isobar). The thermodynamics of the
generalized CS model as described in Sec.~\ref{sec:gcs} is thus ideally suited
in the role of benchmark for the thermodynamics of gases with more generic
statistical or dynamical interactions. From work in progress on several model
systems in $\mathcal{D}$ dimensions with statistical interactions not
restricted as in (\ref{eq:19}), we have compelling evidence that none of the
ideal quantum gas hallmarks (i)--(iii) are upheld anymore. The stage is thus
set for the exact analysis of qualitatively new thermodynamic phenomena
including phase transitions in non-bosonic quantum gases.

%\pagebreak
%%%%%%%%%%%%%%%%%%%%%%%%%%%%%%%%%%%%%%%%%%%%%%%
%
\acknowledgments
%
%%%%%%%%%%%%%%%%%%%%%%%%%%%%%%%%%%%%%%%%%%%%%%%
Financial support from the DFG Schwerpunkt \textit{Kollektive Quantenzust{\"a}nde
  in elektronischen 1D {\"U}bergangsmetallverbindungen} (for M.K.)  is gratefully
acknowledged. G.M. is grateful to Prof. Dr. H. Thomas for stimulating discussions.

%%%%%%%%%%%%%%%%%%%%%%%%%%%%%%%%%%%%%%%%%%%%%%%%
%\begin{thebibliography}{100}
%\section*{References}    

%{\bibliography{../../../refnew,../../../muller,../../../references}}%
%{\bibliography{/home/karbach/REFERENCES/references}}
%\bibliographystyle{jpa}
%\end{thebibliography}
%%%%%%%%%%%%%%%%%%%%%%%%%%%%%%%%%%%%%%%%%%%%%%%%
\end{document}